\documentclass[10pt]{iopart}

\usepackage{graphicx}
\usepackage{dcolumn}
\usepackage{bm}

\usepackage{color}
\usepackage{amssymb}
\usepackage[implicit=true,colorlinks=true,linkcolor=blue,
citecolor=blue,urlcolor=blue]{hyperref} 
\begin{document}

\title[Magnetic-field orientation dependence]{Magnetic-field orientation dependence of transport properties of topologically superconducting wire}


\author{V~V~Val'kov and S~V~Aksenov}

\address{%
  Laboratory of Theoretical Physics, Kirensky Institute of Physics, Federal Research Center KSC SB RAS, Akademgorodok street 50/38, Krasnoyarsk, Russia}%
\ead{vvv@iph.krasn.ru,asv86@iph.krasn.ru}

\begin{abstract}
We present the study of transport properties of a superconducting wire with strong Rashba spin-orbit coupling for different orientations of an external magnetic field. Using the nonequilibrium Green's functions in the tight-binding approach the crucial impact of the relative alignment of lead magnetization and the Majorana bound state (MBS) spin polarization on the low-bias conductance and shot noise is presented. Depending on this factor the transport regime can effectively vary from symmetric to extremely asymmetric. In the last situation the current-symmetry breaking, the suppression of the MBS-assisted conductance and specific Fano factor behavior lead to current-switch effect. The persistence of these features under the presence of diagonal disorder and phenomenologically modeled g-factor anisotropy is demonstrated. In the case of paramagnetic leads the MBS spin polarization gives rise to the spin-filtering effect depending on the magnetic-field orientation.
\end{abstract}
\maketitle
\ioptwocol


\section{\label{sec1}Introduction}

The pursuit of experimental observation of a Majorana fermion originally started in the field of particle physics continued then in the solid-state systems \cite{wilczek-09,elliott-15}. Within the framework of the Kitaev model \cite{kitaev-01} it was shown that the emergent quasiparticles appearing in the many-body system with superconducting (SC) pairing possess the self-Hermitian property of the Majorana fermions. Additionally, these Majorana bound states (MBSs) induced in low-dimensional structures have exponentially small or zero energy and are spatially separated. The last feature opens a way to utilize the MBSs as the building blocks of fault-tolerant quantum computers \cite{kitaev-03}. Since the MBSs obey non-Abelian statistics \cite{ivanov-01} quantum information encoded in an MBS-based qubit can be manipulated by braiding operations \cite{nayak-08,alicea-11}.

The existence of the MBSs was predicted in many systems such as the surface of superfluid $^3$He-B \cite{volovik-09}, the $\nu=5/2$ fractional quantum Hall system (the Moore-Read Pfaffian state) \cite{moore-91}, chiral p-wave SC \cite{dassarma-06}, the edge of 2D- \cite{fu-09} and 3D topological insulator \cite{fu-08,cook-11}. For two 1D structures, the chain of magnetic atoms \cite{nadj-perge-13,pientka-13,nadj-perge-14} and semiconducting wire \cite{mourik-12,das-12,deng-12}, the experimental proofs of the presence of the MBSs were provided. In last case the InSb or InAs wires characterized by strong spin-orbit coupling (SOC) in proximity to an s-wave SC were investigated. Under the influence of an external magnetic field an effective p-wave pairing is realized in the wire and two MBSs appear at its opposite ends \cite{lutchyn-10,oreg-10}. The tunneling spectroscopy measurements reveal the zero-bias conductance peak (ZBP) indicating resonant Andreev transport processes through the zero-energy MBS \cite{flensberg-10,wimmer-11,wu-12}. Among with this feature there are a few different, e.g. the fractional Josephson effect \cite{fu-09,lutchyn-10,ioselevich-11}, electron teleportation \cite{bolech-07,fu-10} and the peculiarities of current fluctuations \cite{nillson-08,law-09}, intensively studied in topologically SC systems in the last decade. These consequences of the self-Hermitian property of the MBSs can be used both for their detection and in different applications.

\begin{figure}[tb]
	\includegraphics[width=0.45\textwidth]{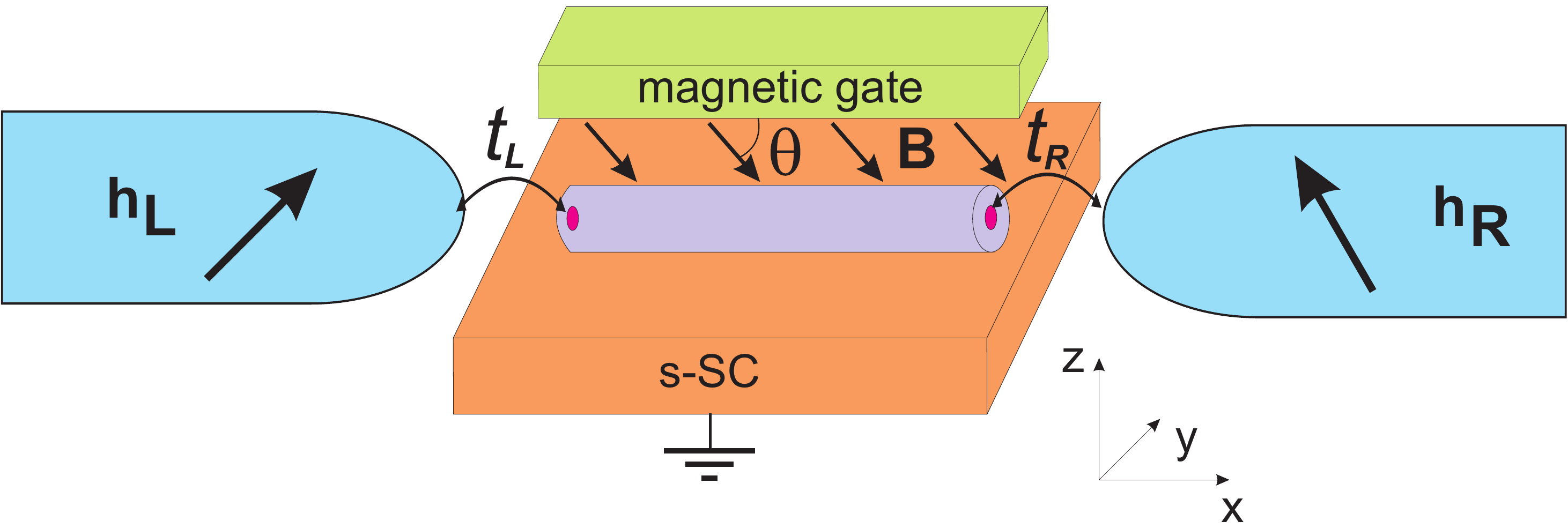}
	\caption{\label{1} The superconducting wire between ferromagnetic leads. Red circles indicate MBSs.}
\end{figure}
It is essential that the Majorana nature of the ZBP is not a single interpretation. It was shown that disorder \cite{bagrets-12,pikulin-12}, size quantization \cite{kells-12,liu-12} and the Kondo correlations \cite{lee-12} can also lead to the ZBP. Therefore, it is necessary to analyze supplementary information concerning quantum transport via the MBSs. One of the important characteristics is the spin polarization of the MBS \cite{sticlet-12,nagai-14}. It was demonstrated in \cite{sticlet-12} that this value is significantly nonzero at the edges of the wire and decays exponentially into the center. Additionally, the sign of the spin polarization at the edges depends on the type of the SOC \cite{sticlet-12} and the external magnetic-field orientation \cite{valkov-16a,valkov-17b}. The Andreev reflection (AR) \cite{andreev-64} is also nontrivially affected by the MBS spin polarization leading to the selective equal-spin \cite{he-14,law-14} and noncollinear AR \cite{wu-14}. Thus, this characteristic can be considered as a local order parameter of topological phase and probed in spin-polarized scanning tunneling spectroscopy/microscopy measurements to detect the MBSs.

Taking into account these properties in the present article we analyze the influence of magnetic-field orientation on quantum transport via the MBSs. This problem was already partly investigated in the experiment \cite{mourik-12} with spin-degenerate transport channels and revealed the presence of the ZBP for all magnetic-field angles in the plane perpendicular to the Rashba effective field. The importance of this issue is additionally emphasized by the recent experiments uncovered the g-factor anisotropy of low-dimensional semiconducting structures \cite{lee-14,qu-16}.
Here studying spin-degenerate and -polarized transport we show that the low-bias conductance brings the features allowing both to detect the MBSs and employ that unusual behavior in electronic applications, such as a current switch and spin filter. In work \cite{dessotti-16} the current switch assisted by the MBSs was already studied. The effect was induced by the combination of quantum phase transition and destructive interference. In contrast, our proposal is based on the magnetic-gate control of the MBS spin polarization as it is schematically shown in Fig. \ref{1}. It is also demonstrated that the found transport features persist in case of diagonal disorder and phenomenologically modeled g-factor anisotropy.

The article has been organized in some sections. The model Hamiltonian is described in Section \ref{sec2}. The derivation of current formula in terms of the nonequilibrium Green's functions in the tight-binding approach is discussed in Section \ref{sec3}. The features of the MBS spin polarization are analyzed in Section \ref{sec4}. The magnetic-field orientation dependence of the conductance and shot noise is considered in Section \ref{sec5}. In Section \ref{sec6} we discuss the influence of Anderson disorder and g-factor anisotropy on the conductance and noise. Conclusions are given in the Section \ref{sec7}.

\section{\label{sec2}The model Hamiltonian}

Let us consider a semiconducting wire with the strong Rashba SOC deposited on a grounded s-type SC substrate (see Fig. \ref{1}). Hereinafter we name it a 'superconducting wire' taking into account an induced SC pairing in the wire characterized by a parameter $\Delta$. An external magnetic field $\mathbf{B}=\left(B_{x},0,-B_{z}\right)=B\left(\cos\theta,0,-\sin\theta\right)$ is assumed to be oriented at an arbitrary angle in the plane perpendicular to the Rashba effective field $\mathbf{B}_{SO}~\parallel~y$. We suppose that in electronic applications the direction and amplitude of the magnetic field can be manipulated by a 'magnetic gate', e.g. as it was proposed to manipulate domain-wall motion in \cite{murapaka-16}.

The SC wire is described by the following microscopic Hamiltonian:
 \begin{eqnarray} \label{HW}
&&\hat{H}_{W} =\sum\limits_{j=1}^{N}\left[\sum\limits_{\sigma}\xi_{\sigma}a^+_{j\sigma}a_{j\sigma}+
\Delta a_{j\uparrow}a_{j\downarrow} - V_{x} a^+_{j\uparrow}a_{j\downarrow} + H.c.\right]\nonumber\\
&&~~~~~~~-\sum\limits_{\sigma;j=1}^{N-1}\left[\frac{t}{2}a^+_{j\sigma}a_{j+1,\sigma}+\frac{\alpha}{2}\sigma a^+_{j\sigma}a_{j+1,\overline{\sigma}}+H.c.\right],~~
\end{eqnarray}
where $a_{j\sigma}$ - an electron annihilation operator on $j$th site of the wire with spin $\sigma$; $\xi_{\sigma}=t+\sigma V_{z}-\mu$ - an on-site energy of the electron with spin $\sigma$ taking into account the Zeeman component $V_{z}$; $\mu$ - a chemical potential of the system; $V_{x\left(z\right)}=\mu_B B_{x\left(z\right)}$ - $x$- and $z$-components of the Zeeman energy; $t$ - a nearest neighbor hopping parameter; $\alpha$ - an intensity of the Rashba SOC. From experimental point of view it is interesting to note that the longitudinal component of the magnetic field can be treated as an intrinsic SOC \cite{kwapinski-17}. In turn that is possible to control $V_{x}$ by gate voltage by analogy with the extrinsic (Rashba) SOC \cite{nitta-97}.

The wire on the substrate is situated between the ferromagnetic leads. The corresponding mean-field Hamiltonians are given by
\begin{eqnarray} \label{HLR}
&&\hat{H}_{i} =\sum\limits_{l\sigma}\biggl\{\left[\xi_{l}-\frac{eV_{i}}{2}-\sigma M_{i} \cos\theta_{i} \right]c^+_{l\sigma}c_{l\sigma}~~~\biggr.\nonumber \\
&&~~~~~~~~~~~~~~~~~~~~~~~~~~\biggl.-M_{i}\sin\theta_{i} c^+_{l\sigma}c_{l\overline{\sigma}}\biggr\},~i=L,R,
\end{eqnarray}
where $c^+_{l\sigma}$ - an electron creation operator in $i$th lead with a wave vector $l$, spin $\sigma$ and an energy $\xi_{l}=\epsilon_{l}-\mu$; $M_{i}=\frac{1}{2} g \mu_{B} h_{i}$ - an energy of the $i$th lead magnetization $\mathbf{h}_{i}$; $\theta_{i}$ - an angle between $\mathbf{h}_{i}$ and $z$ axis in the $xz$ plane; $\sigma=\pm1$ or $\uparrow,\downarrow$. The bias voltage $\pm V/2$ is applied to the left (right) lead.

The interaction between the leads and the SC wire is given by a standard tunnel Hamiltonian,
\begin{equation} \label{HT}
\hat{H}_{T} =t_{L}\sum \limits_{k\sigma}c_{k\sigma}^{+} a_{1\sigma} +
t_{R}\sum \limits_{p\sigma} c_{p\sigma}^{+} a_{N\sigma} + H.c.,
\end{equation}
where $t_{L\left(R\right)}$ - a tunnel parameter between the left (right) lead and the wire. Thus the total Hamiltonian of the system is $\hat{H}=\hat{H}_{L}+\hat{H}_{R}+\hat{H}_{T}+\hat{H}_{W}$.

\section{\label{sec3}Current in terms of the nonequilibrium Green's functions in the tight binding approach}

To calculate the spin-dependent transport properties of the SC wire modeled by the microscopic tight-binding-type Hamiltonian (\ref{HW}) we employ the nonequilibrium Green's functions \cite{keldysh-65,datta-95,datta-05} in the spin$\otimes$Nambu space \cite{wu-12,wu-14}.

In order to simplify the derivation of the current formula we make a few diagonalizing steps before. Firstly, it is the Bogolubov transformation for the ferromagnetic contacts \cite{zeng-03},
\begin{equation} \label{Bog1}
c_{k\left(p\right)\sigma}=\alpha_{k\left(p\right)\sigma}\cos\frac{\theta_{L\left(R\right)}}{2}
-\sigma\alpha_{k\left(p\right)\overline{\sigma}}\sin\frac{\theta_{L\left(R\right)}}{2}.
\end{equation}
Secondly, we introduce the field operators in the spin$\otimes$Nambu space which allow to take into account both the SC pairing and spin-flip processes in the system,
\begin{eqnarray} \label{Nambu}
&&\widehat{\psi}_{k\left(p\right)}=\left(\alpha_{k\left(p\right)\uparrow}~\alpha_{k\left(p\right)\downarrow}^{+}~ \alpha_{k\left(p\right)\downarrow}~\alpha_{k\left(p\right)\uparrow}^{+}\right)^T,~\\
&&\widehat{\psi}_{W}=\left(a_{1\uparrow}~a_{1\downarrow}^{+}~a_{1\downarrow}~a_{1\uparrow}^{+}~...~
a_{N\uparrow}~a_{N\downarrow}^{+}~a_{N\downarrow}~a_{N\uparrow}^{+}\right)^T.\nonumber
\end{eqnarray}
As a result the summands in $\hat{H}$ become
\begin{eqnarray} \label{HWT1}
&&\hat{H}_W = \widehat{\psi}_{W}^{+}\widehat{h}_{W}\widehat{\psi}_{W}+const,~\\
&&\hat{H}_T =\sum\limits_{k}\widehat{\psi}_{k}^{+}\widehat{T}_{L}\left(t\right)\widehat{P}_{1}\widehat{\psi}_{W}+
\sum\limits_{p}\widehat{\psi}_{p}^{+}\widehat{T}_{R}\left(t\right)\widehat{P}_{N}\widehat{\psi}_{W}+H.c.,\nonumber
\end{eqnarray}
where $\widehat{h}_{W}$ - an $4N \times 4N$ matrix with the following nonzero blocks
\begin{eqnarray} \label{Hj}
&&H_{jj}=
\frac{1}{2}\left(\begin{array}{cccc}
\xi_{\uparrow} & \Delta & -V_{x} & 0 \\
\Delta & -\xi_{\downarrow} & 0 & V_{x} \\
-V_{x} & 0 & \xi_{\downarrow} & -\Delta \\
0 & V_{x} & -\Delta & -\xi_{\uparrow}
\end{array}\right),~\\
&&H_{j,j+1}=H_{j,j-1}=
\frac{1}{2}\left(\begin{array}{cccc}
-\frac{t}{2} & 0 & -\frac{\alpha}{2} & 0 \\
0 & \frac{t}{2} & 0 & -\frac{\alpha}{2} \\
\frac{\alpha}{2} & 0 & -\frac{t}{2} & 0 \\
0 & \frac{\alpha}{2} & 0 & \frac{t}{2}
\end{array}\right).\nonumber
\end{eqnarray}
The tunnel coupling matrices, $\widehat{T}_{i}\left(t\right)$ ($i=L,R$), are received by performing a gauge transformation \cite{zeng-03,rogovin-74},
\begin{eqnarray} \label{TLR}
&&\widehat{T}_{i}\left(t\right)=\widehat{R}_{i}\widehat{t}_{i}\left(t\right),\\
&&\widehat{R}_{i}=\left(\begin{array}{cccc}
\cos\frac{\theta_{i}}{2} & 0 & \sin\frac{\theta_{i}}{2} & 0 \\
0 & \cos\frac{\theta_{i}}{2} & 0 & -\sin\frac{\theta_{i}}{2} \\
-\sin\frac{\theta_{i}}{2} & 0 & \cos\frac{\theta_{i}}{2} & 0 \\
0 & \sin\frac{\theta_{i}}{2} & 0 & \cos\frac{\theta_{i}}{2}
\end{array}\right),\nonumber\\
&&\widehat{t}_{i}\left(t\right)=
\frac{t_{i}}{2}\left(\begin{array}{cccc}
e^{-i\frac{eV}{2}t} & 0 & 0 & 0 \\
0 & -e^{i\frac{eV}{2}t} & 0 & 0 \\
0 & 0 & e^{-i\frac{eV}{2}t} & 0 \\
0 & 0 & 0 & -e^{i\frac{eV}{2}t}
\end{array}\right),\nonumber
\end{eqnarray}
The $4 \times 4N$ matrices $\widehat{P}_{1}=\left(\widehat{I}~\widehat{O}\right)$ and $\widehat{P}_{N}=\left(\widehat{O}~\widehat{I}\right)$ act as projection operators and consist of a $4 \times 4$ unit matrix, $\widehat{I}$, and a zero block, $\widehat{O}$ \cite{valkov-17a}.

Let us introduce the matrix nonequilibrium Green's functions in terms of the above-described field operators (\ref{Nambu}) as
\begin{eqnarray} \label{GF}
&&\widehat{G}_{nm}\left(\tau,\tau'\right)=-i\left\langle\widehat{T}_{C}\widehat{\psi}_{n}\left(\tau\right)\otimes
\widehat{\psi}_{m}^{+}\left(\tau'\right)\right\rangle,\\
&&~~~~~~~~~~~~~~~~~~~~~~~~~~~~~~~~~~~~~~~~~~~~~~~~n,m=k,p,W,\nonumber
\end{eqnarray}
where $\widehat{T}_{C}$ is a Keldysh-contour ordering operator. Next the current in the left lead is given by $\widehat{I}_{L}=e\dot{N}_{L}$ ($N_{L}=\sum_{k\sigma}\alpha_{k\sigma}^{+}\alpha_{k\sigma}$ is a particle operator in the left lead). And, after some manipulations, we finally find
\begin{eqnarray} \label{IL1}
&&\langle\widehat{I}_{L}\rangle=2e\sum_{k}Tr\Biggl[Re\Bigl\{\widehat{\sigma}\widehat{T}_{L}^{+}\left(t\right)
\widehat{G}_{kW}^{<}\left(t,~t\right)\widehat{P}_{1}^{+}\Bigr\}\Biggr]=\nonumber\\
&&=e\int\limits_{-\infty}^{+\infty}\frac{d\omega}{\pi}Tr\Biggl[Re\Bigl\{\widehat{\sigma}
\left(\widehat{\Sigma}_{L}^{r}\widehat{G}_{1,1}^{<}+
\widehat{\Sigma}_{L}^{<}\widehat{G}_{1,1}^{a}\right)\Bigr\}\Biggr],
\end{eqnarray}
where $\widehat{\sigma}=diag\left(1,-1,1,-1\right)$ accounts for the electron and hole transport channels; $\widehat{\Sigma}_{L}\left(t,t'\right)=\widehat{T}_{L}^{+}\left(t\right)\widehat{g}_{k}\left(t,~t'\right)\widehat{T}_{L}\left(t'\right)$ - the matrix self-energy function describing the influence of the left lead on the wire; $\widehat{g}_{k}\left(t,~t'\right)$ - the free-particle Green's function of the left lead; $\widehat{G}_{j,j}^{<,a}=\widehat{P}_{j}\widehat{G}_{W}^{<,a}\widehat{P}_{j}^{+}$ - the Fourier transforms of the lesser and advanced Green's functions of the wire which are projected on the $j$th site subspace.

The retarded and lesser/greater Green's functions can be obtained from the Dyson and Keldysh equations, respectively,
\begin{eqnarray}
&&\widehat{G}_{W}^{r}=\left(\omega-\widehat{h}_{W}-\widehat{\Sigma}^{r}\right)^{-1},~
\widehat{G}_{W}^{a}=\left(\widehat{G}_{W}^{r}\right)^{+},~\label{Gr}\\
&&\widehat{G}_{W}^{\lessgtr}=\widehat{G}_{W}^{r}\widehat{\Sigma}^{\lessgtr}\widehat{G}_{W}^{a}.~\label{G+-}
\end{eqnarray}
In expressions (\ref{Gr}), (\ref{G+-}) the total self-energy function of the system is $\widehat{\Sigma}^{n}=\widehat{P}_{1}^{+}\widehat{\Sigma}_{L}^{n}\widehat{P}_{1}+
\widehat{P}_{N}^{+}\widehat{\Sigma}_{R}^{n}\widehat{P}_{N}$ and the $i$th lead components are $\widehat{\Sigma}_{i}^{r}=-\frac{i}{2}\widehat{\Gamma}_{i}$, $\widehat{\Sigma}_{i}^{<}=\left(\widehat{\Sigma}_{i}^{a}-\widehat{\Sigma}_{i}^{r}\right)
\widehat{F}_{i}$, where
\begin{eqnarray} \label{GLR}
&&\widehat{\Gamma}_{i}=\left(\begin{array}{cccc}
\Gamma_{i11} & 0 & \Gamma_{i12} & 0 \\
0 & \Gamma_{i22} & 0 & \Gamma_{i12} \\
\Gamma_{i12} & 0 & \Gamma_{i22} & 0 \\
0 & \Gamma_{i12} & 0 & \Gamma_{i11}
\end{array}\right),\\
&&\widehat{F}_{L\left(R\right)}=diag\biggl(n\left(\omega \pm eV/2\right),~n\left(\omega \mp eV/2\right),\biggr.\nonumber\\
&&\biggl.~~~~~~~~~~~~~~~~~~~~~~~~~~n\left(\omega \pm eV/2\right),~n\left(\omega \mp eV/2\right)\biggr).\nonumber
\end{eqnarray}
Here $\Gamma_{i11}=\Gamma_{i}^{+}\cos^2\frac{\theta_{i}}{2}+\Gamma_{i}^{-}\sin^2\frac{\theta_{i}}{2}$, $\Gamma_{i22}=\Gamma_{i}^{+}\sin^2\frac{\theta_{i}}{2}+\Gamma_{i}^{-}\cos^2\frac{\theta_{i}}{2}$, $\Gamma_{i12}=\frac{1}{2}\left(\Gamma_{i}^{+}e^{-i\phi_{i}}-\Gamma_{i}^{-}e^{i\phi_{i}}\right)\sin\theta_{i}$. $\Gamma_{i}^{+\left(-\right)}=2\pi\left(\frac{t_{i}}{2}\right)^2\rho_{i}^{+\left(-\right)}$ - the coupling strength between the wire and the majority (minority) subband of the $i$th lead; $\rho_{i}^{+\left(-\right)}$ - the DOS of the majority (minority) subband of the $i$th lead; $\phi_{i}$ - an $i$th lead magnetization angle in the $xy$ plane. In further analysis we use the convention: $\phi_{i}=0$ if $\theta_{i}>0$ and $\phi_{i}=\pi$ if $\theta_{i}<0$ \cite{wu-14}. In the calculations below the leads are treated in the wide-band limit. It means that $\Gamma_{i}^{+\left(-\right)}=const$. The $i$th lead polarization, $P_{i}$, defines the degree of the spin polarization of the carriers, $P_{i}=\left(\Gamma_{i}^{+}-\Gamma_{i}^{-}\right)/\left(\Gamma_{i}^{+}+\Gamma_{i}^{-}\right)$. In this study, unless stated otherwise, we focus on the transport properties of the half-metallic leads, $P_{L}=P_{R}=P=1$ (NiMnSb or CrO$_2$ \cite{keizer-06}) and consider the symmetric transport regime, $\Gamma_{L}=\Gamma_{R}$.

\begin{figure*}[htbp]
	\includegraphics[width=0.425\textwidth]{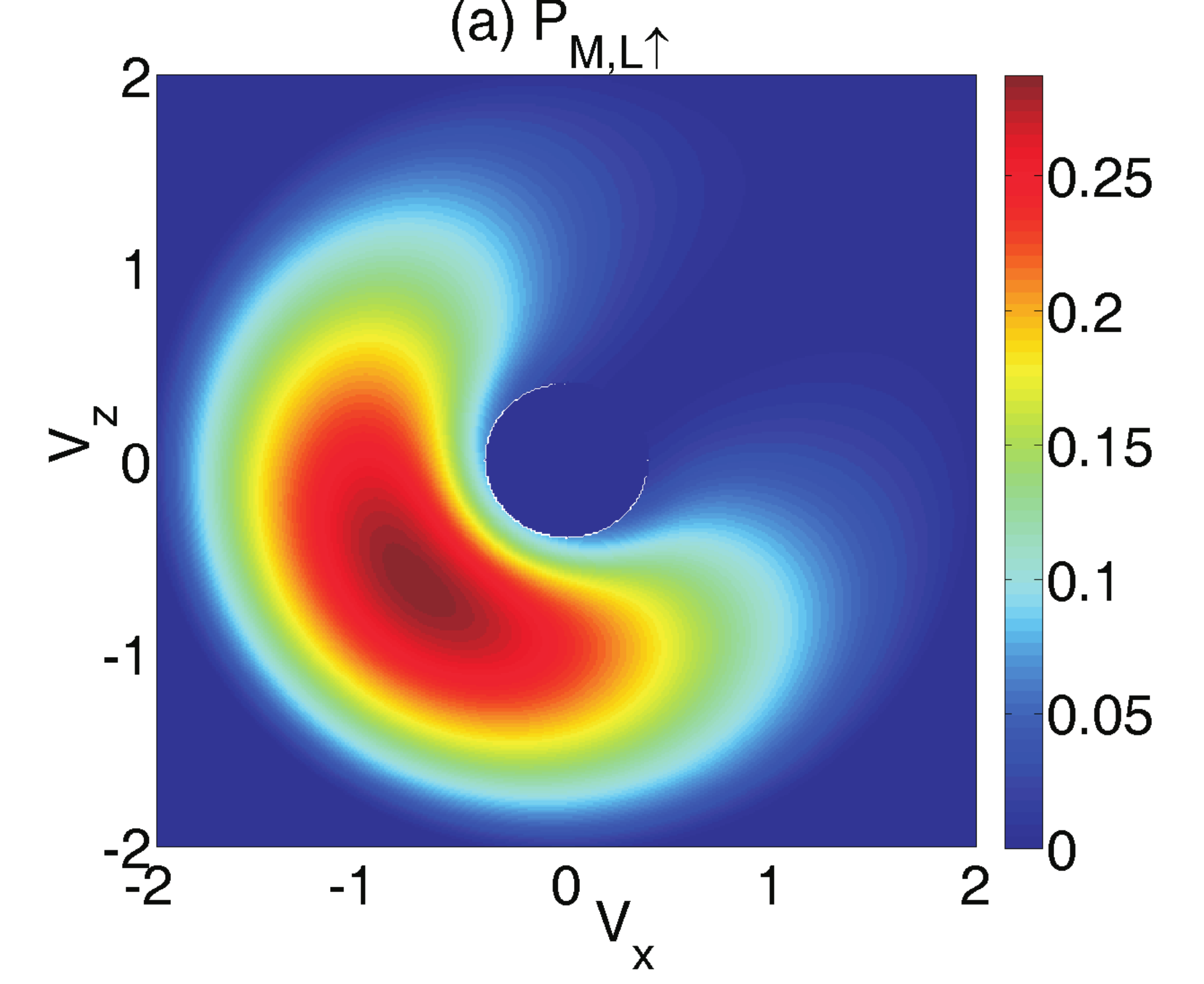}
	\includegraphics[width=0.4\textwidth]{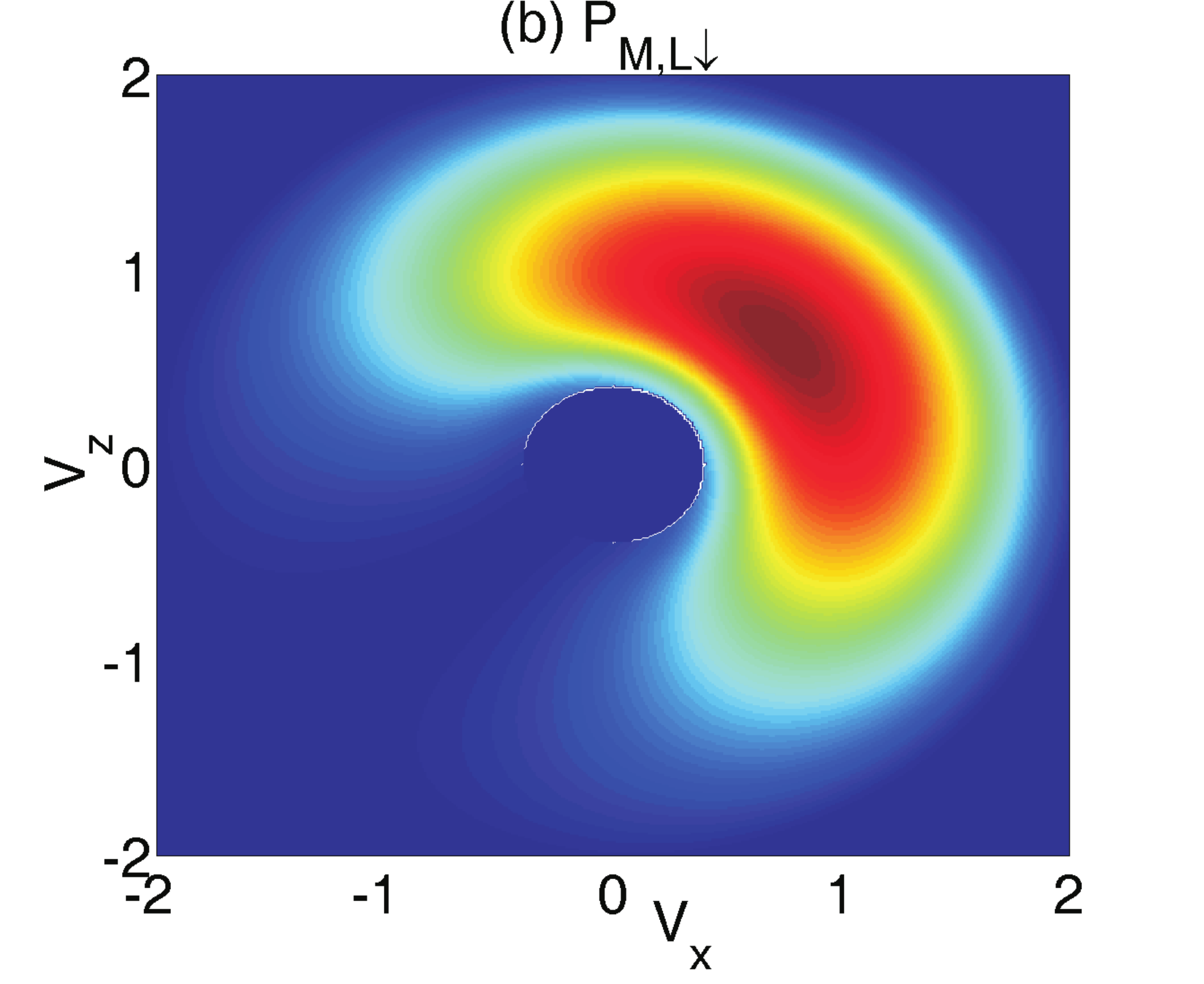}
	\includegraphics[width=0.4\textwidth]{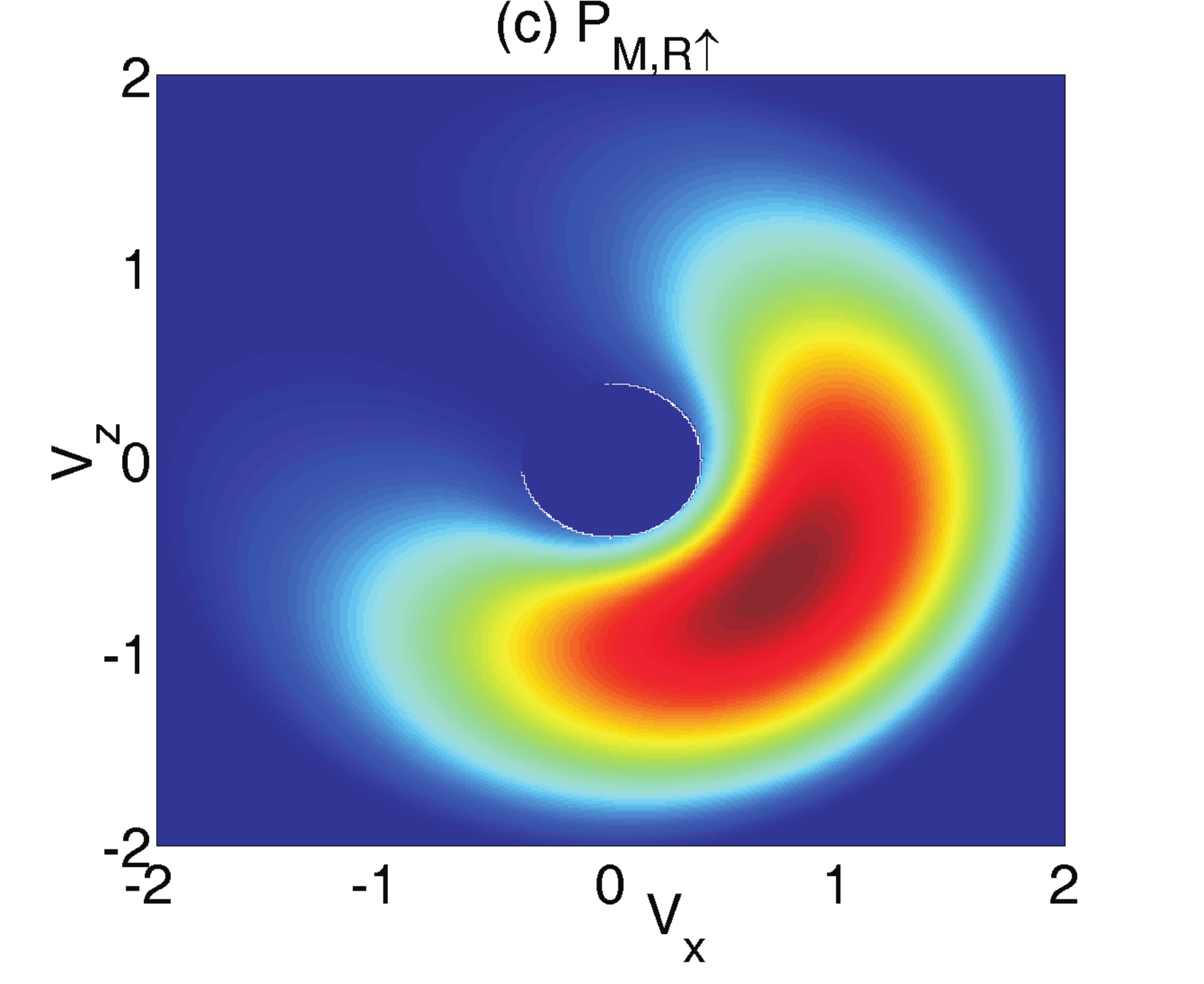}
	\includegraphics[width=0.4\textwidth]{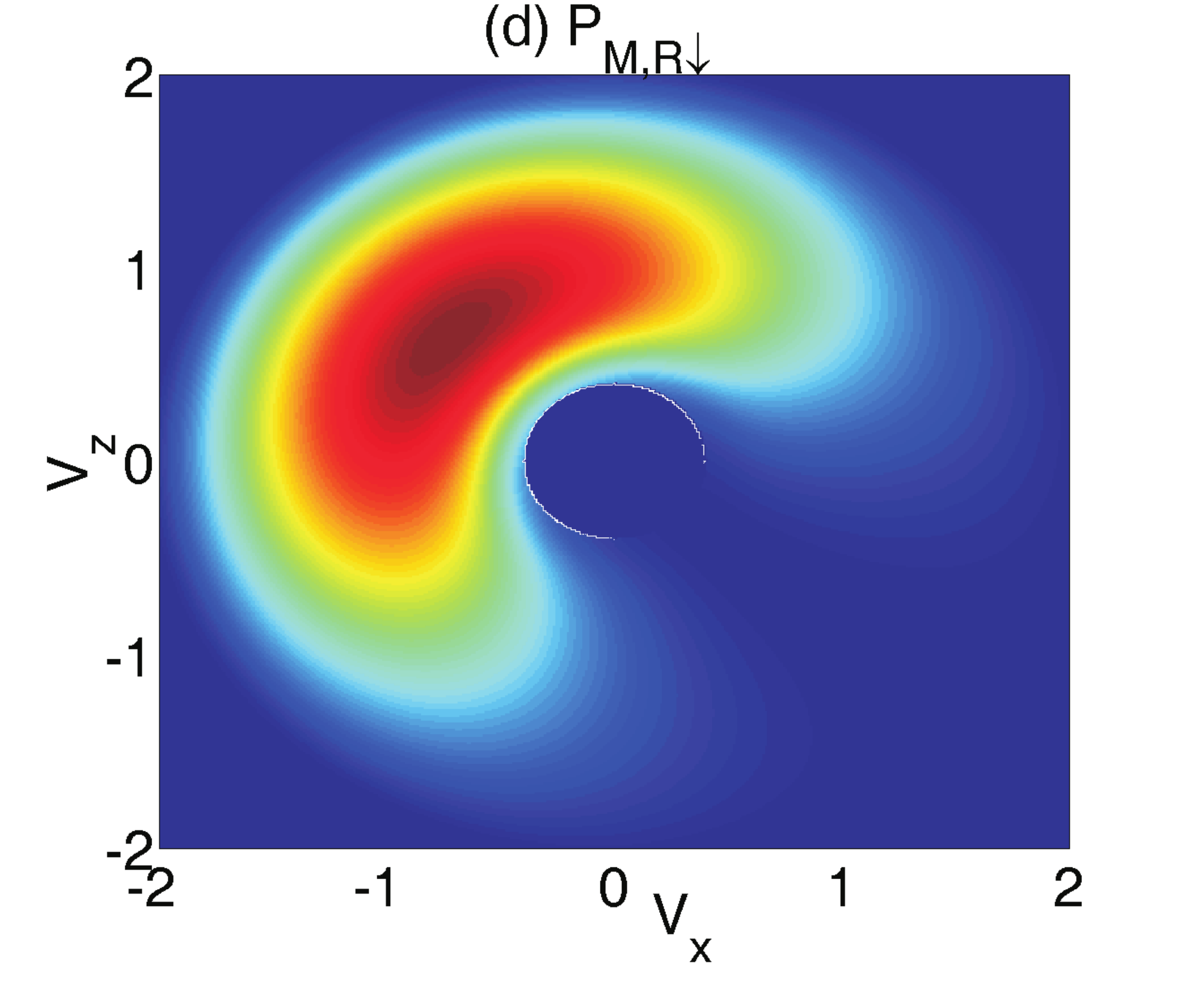}
	\caption{\label{2} The magnetic-field orientation dependence of the MBS spin-dependent probability densities. Parameters: $\mu=0$, $\Delta=0.4$, $\alpha=2$, $N=30$.}
\end{figure*}
As it was already shown before the nonlocality of the MBSs and their coupling can be effectively probed by the measurements of the current fluctuations, $\delta I\left(t\right)=I\left(t\right)-\langle I\left(t\right)\rangle$ \cite{bolech-07,nillson-08,law-09,wu-14}. In this work we analyze the autocorrelations of the current in the leads. In particular, the noise spectral density in the left lead can be written as
\begin{equation} \label{Sgen}
S_{L}\left(\omega\right)=\int dt e^{i\omega t} \langle\delta I_L\left(t\right) \delta I_L\left(0\right) + \delta I_L\left(0\right) \delta I_L\left(t\right)\rangle.~
\end{equation}
Substituting (\ref{IL1}) into (\ref{Sgen}) the zero-frequency shot noise in terms of the nonequilibrium Green's functions is given by \cite{wu-12,khlus-87}
\begin{eqnarray} \label{S0}
&&S_{L}\left(\omega\right)=2e^2\int\limits_{-\infty}^{+\infty}\frac{d\omega}{2\pi}Tr\Biggl[
\widehat{\sigma}\widehat{\Sigma}_{L}^{<}\widehat{\sigma}\widehat{G}_{1,1}^{>}+
\widehat{G}_{1,1}^{<}\widehat{\sigma}\widehat{\Sigma}_{L}^{>}\widehat{\sigma}\\
&&-\widehat{\sigma}\left[\widehat{\Sigma}_{L}\widehat{G}_{1,1}\right]^{<}
\widehat{\sigma}\left[\widehat{\Sigma}_{L}\widehat{G}_{1,1}\right]^{>}-
\left[\widehat{G}_{1,1}\widehat{\Sigma}_{L}\right]^{<}\widehat{\sigma}
\left[\widehat{G}_{1,1}\widehat{\Sigma}_{L}\right]^{>}\widehat{\sigma}\nonumber\\
&&+\widehat{\sigma}\left[\widehat{\Sigma}_{L}\widehat{G}_{1,1}\widehat{\Sigma}_{L}\right]^{>}
\widehat{\sigma}\widehat{G}_{1,1}^{<}+
\widehat{G}_{1,1}^{>}\widehat{\sigma}\left[\widehat{\Sigma}_{L}\widehat{G}_{1,1}\widehat{\Sigma}_{L}\right]^{<}
\widehat{\sigma}
\Biggr],\nonumber
\end{eqnarray}
where the following relations of the nonequilibrium Green's functions are used $\left[AB\right]^{\lessgtr}=A^{r}B^{\lessgtr}+A^{\lessgtr}B^{a}$, $\left[ABC\right]^{\lessgtr}=A^{r}B^{r}C^{\lessgtr}C+A^{r}B^{\lessgtr}C^{a}+A^{\lessgtr}B^{a}C^{a}$ \cite{langreth-72}. The similar expressions for the right current and zero-frequency noise power can be easily derived.

In many experimental and theoretical works concerning the topologically SC wires the relation between the parameters of the system is typically $t \gg \alpha,~V_{x,z},~\Delta$ \cite{mourik-12,wu-12}. Here in the calculations below we choose slightly different condition $t \sim \alpha,~V_{x,z},~\Delta$. In the last case the oscillation period of the Majorana zero mode energy increases, in comparison with the former \cite{valkov-16a}. It considerably simplifies the transport calculations. Additionally, for the sake of simplicity, we will analyze the transport at low temperatures $kT=10^{-10}$ and use the hopping parameter $t=1$ in energy units.

\section{\label{sec4} The MBS spin polarization}

\begin{figure*}[htbp]
	\includegraphics[width=0.425\textwidth]{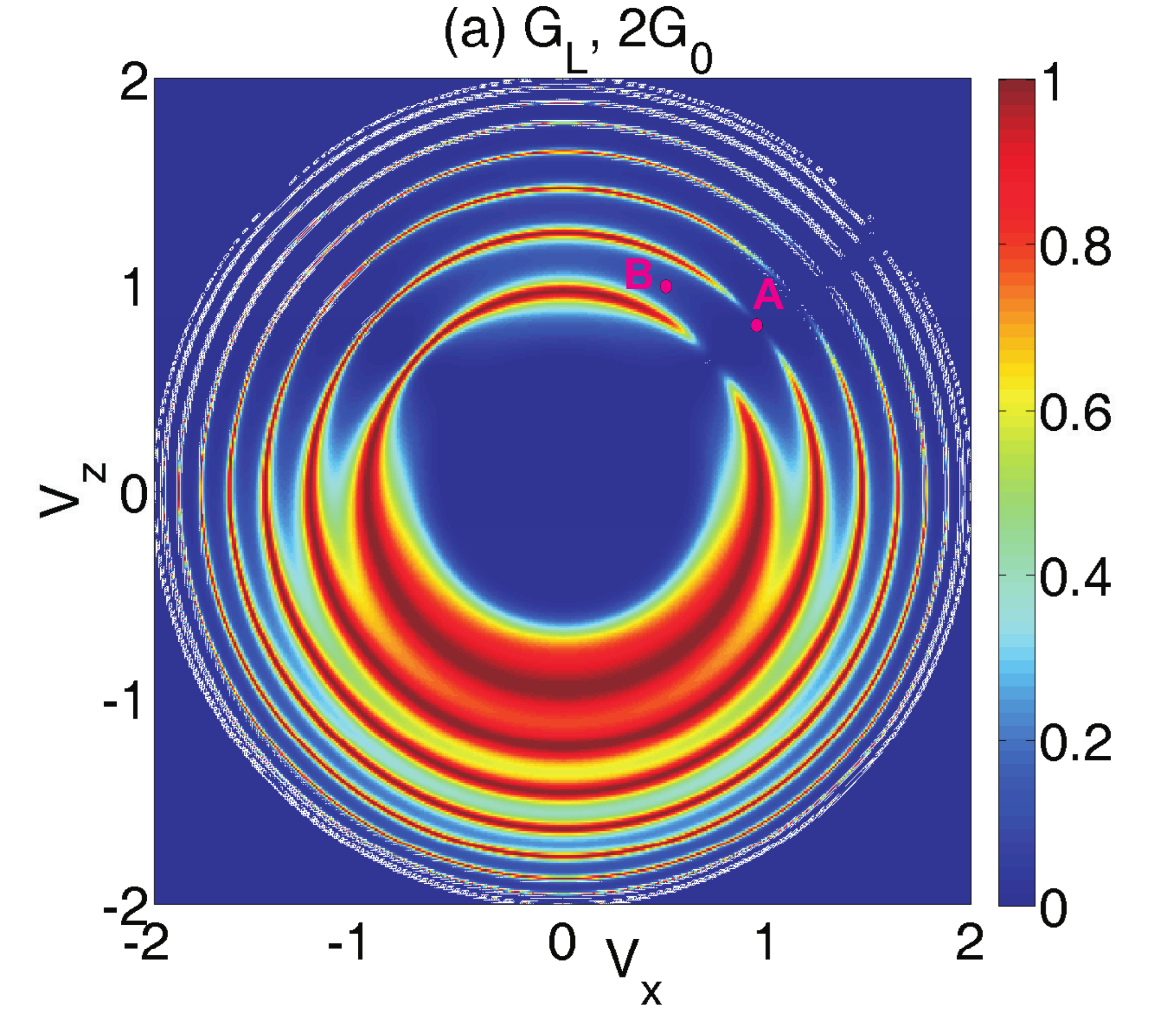}
	\includegraphics[width=0.4\textwidth]{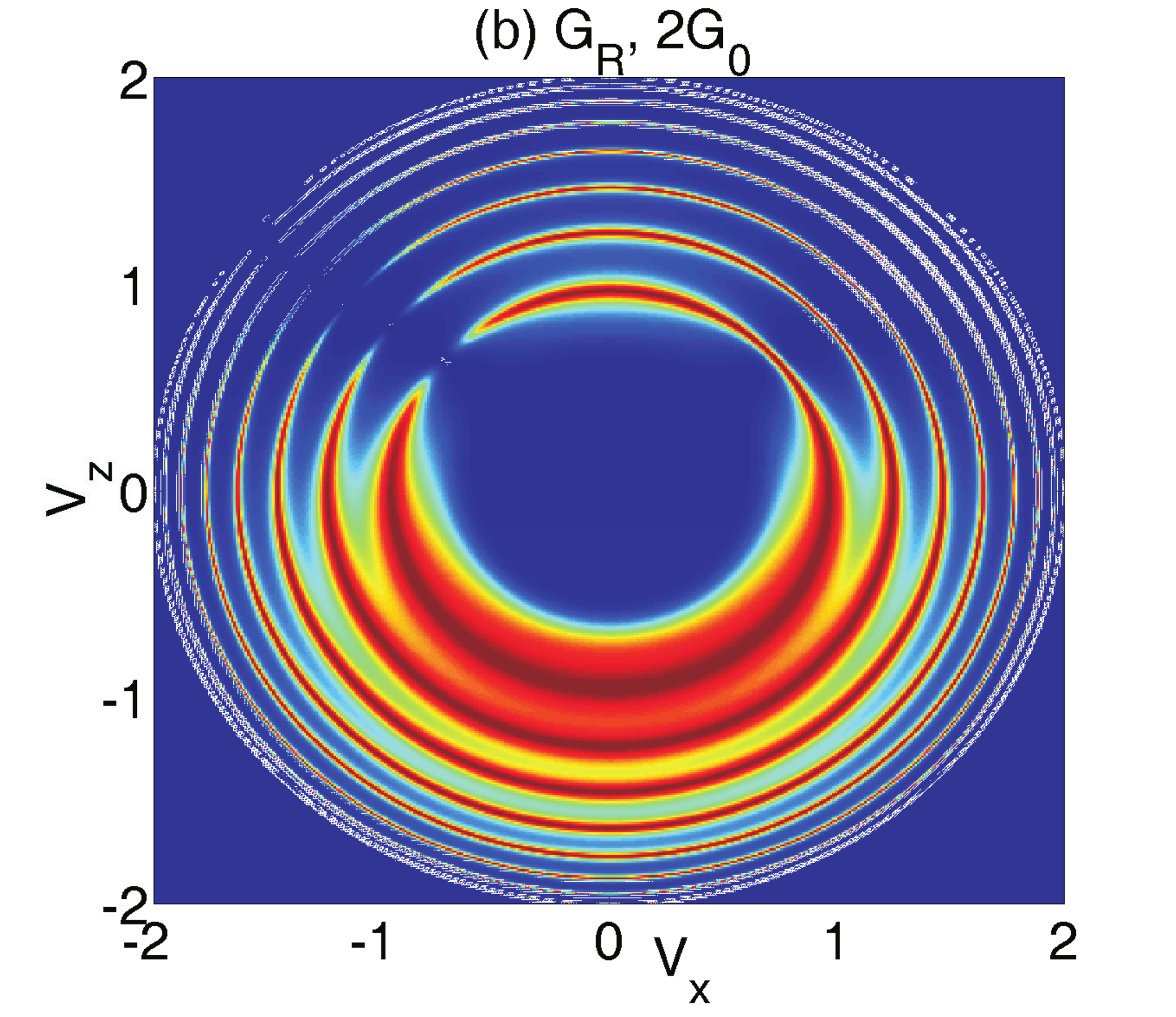}
	\includegraphics[width=0.4\textwidth]{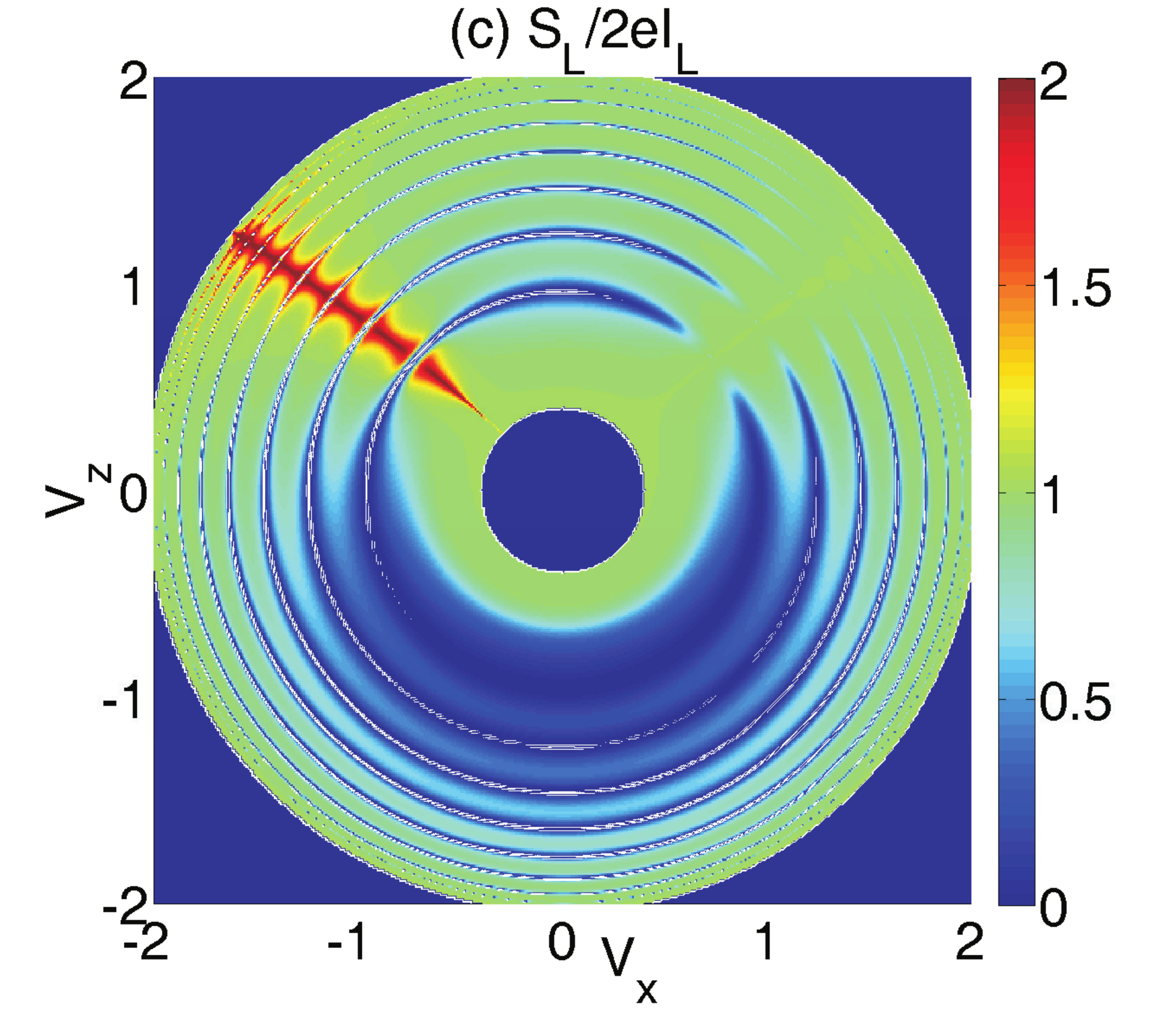}
	\includegraphics[width=0.4\textwidth]{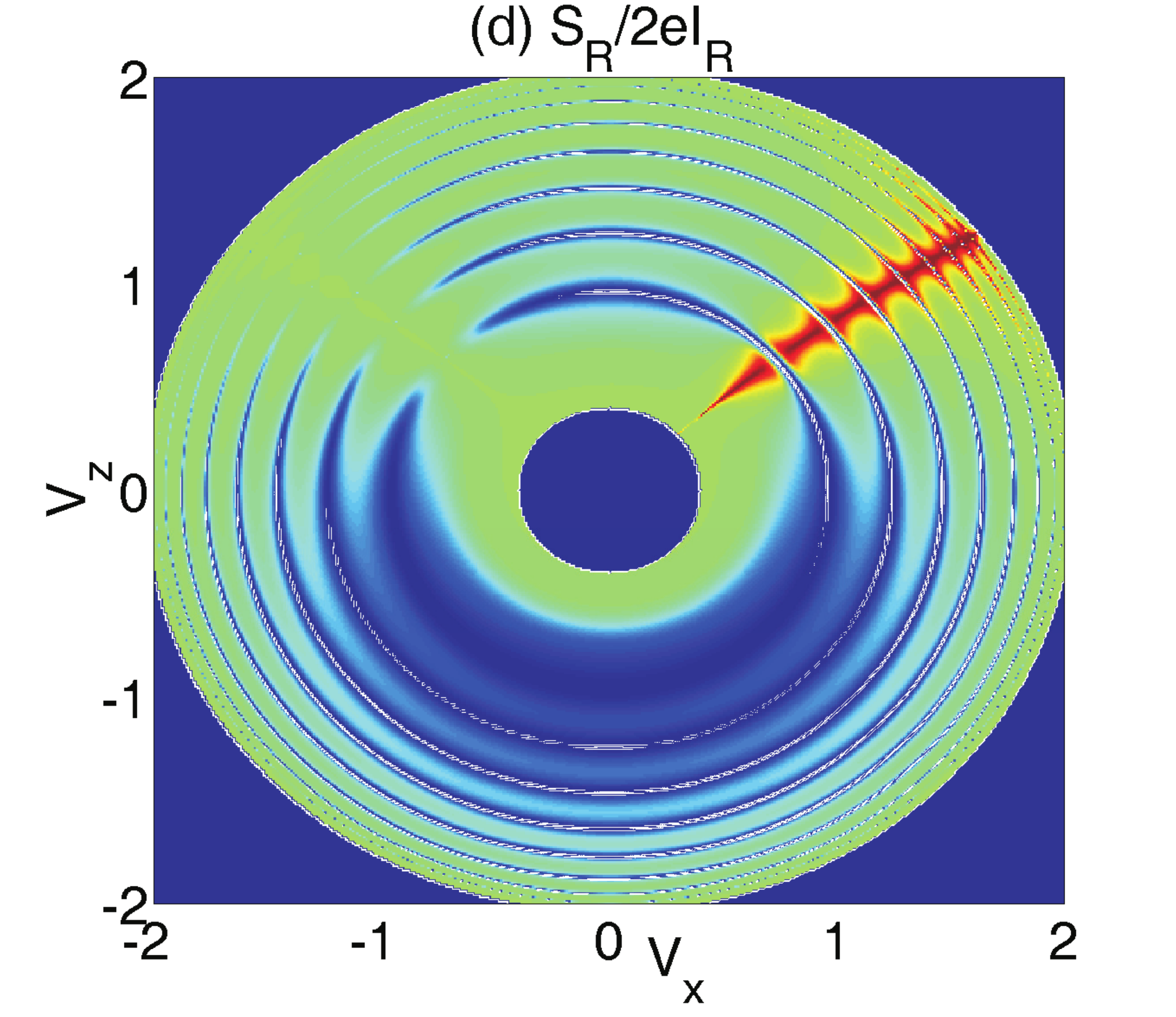}
	\caption{\label{3} The magnetic-field orientation dependence of the low-bias conductance of the left (a) and right (b) lead and the corresponding Fano factors (c, d). Parameters are the same as in Figure \ref{2}.}
\end{figure*}
In the work \cite{sticlet-12} authors showed that the MBS possesses nontrivial spin polarization. It exponentially decreases from the ends to the center of the wire and can have different sign at the edges. In this section we will extend these results to the case of the randomly oriented magnetic field in the $xz$ plane perpendicular to the effective Rashba field. To study the MBS spin polarization we employ the Bogolubov transformation,
\begin{equation} \label{uv}
\beta_l =\sum \limits_{n=1}^{N}\left[u_{ln}a_{n\uparrow}+v_{ln}a_{n\downarrow}^{+}
+w_{ln}a_{n\downarrow}+z_{ln}a_{n\uparrow}^{+}\right].
\end{equation}
In terms of (\ref{uv}) the MBS is treated as an excitation with the lowest positive energy, $E_{M}$, if the parameters of the system satisfy the topological phase conditions, $\mu^2+\Delta^2<V_{x}^2+V_{z}^2<\left(2t - \mu\right)^2+\Delta^2$ \cite{lutchyn-10,oreg-10,valkov-17b}. In order to simplify numerical calculations in the article out of this parametric area all physical quantities are set equal to zero owing to the fact that the spin polarization of the lowest-energy excitation is vanishes and $E_{M}$ quickly increases in topogically trivial phase.

In Figures \ref{2} the spin-resolved MBS probability densities at the left (right) end site of the wire, $P_{M,L\left(R\right)\sigma}$, are demonstrated as functions of the magnetic-field orientation and amplitude. According to the wire's Hamiltonian (\ref{HW}), $P_{M,L\left(R\right)\uparrow}=2\mid u_{M,1\left(N\right)} \mid^2$ ($\mid u_{M,1\left(N\right)} \mid\simeq\mid z_{M,1\left(N\right)} \mid$) have maxima at the lower part of the maps and are suppressed in the opposite fields (see Figs. \ref{2}a,c). The behavior of $P_{M,L\left(R\right)\downarrow}=2\mid v_{M,1\left(N\right)} \mid^2$ ($\mid v_{M,1\left(N\right)} \mid\simeq\mid w_{M,1\left(N\right)} \mid$) is vice versa (see Figs. \ref{2}b,d). Thus, as expected, the maxima for opposite spins at the same edge take place at the fields having opposite sign. Simultaneously, the maxima of $P_{M,i\sigma}$ at opposite edges for given $V_{z}$ appear at the opposite longitudinal fields (compare, for example, Figs. \ref{2}a and c). As we will show further the features of $P_{M,L\left(R\right)\sigma}$ mostly define the MBS-assisted spin-polarized transport in the system.

\begin{figure*}[htbp]
	\includegraphics[width=0.425\textwidth]{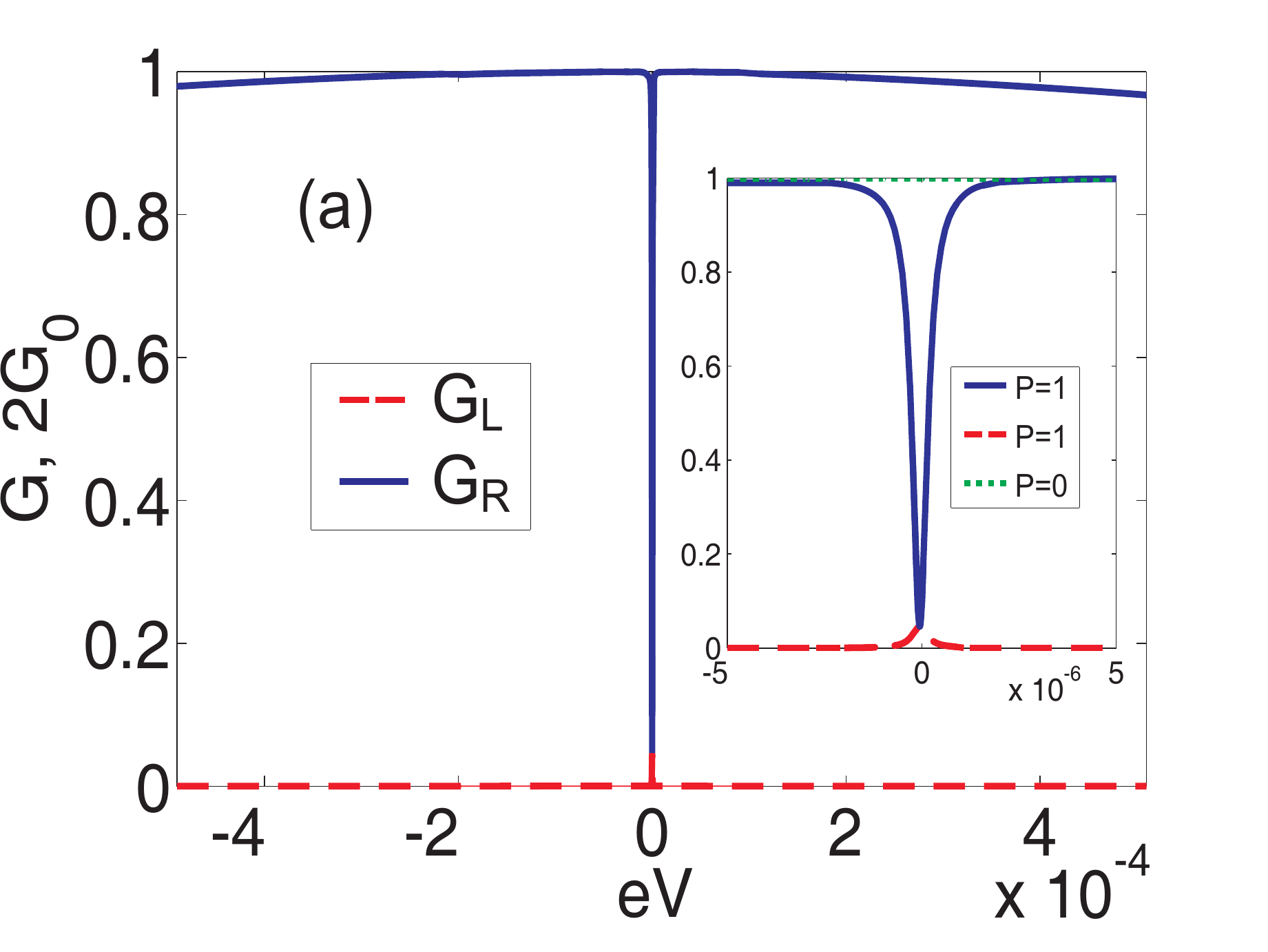}
	\includegraphics[width=0.4\textwidth]{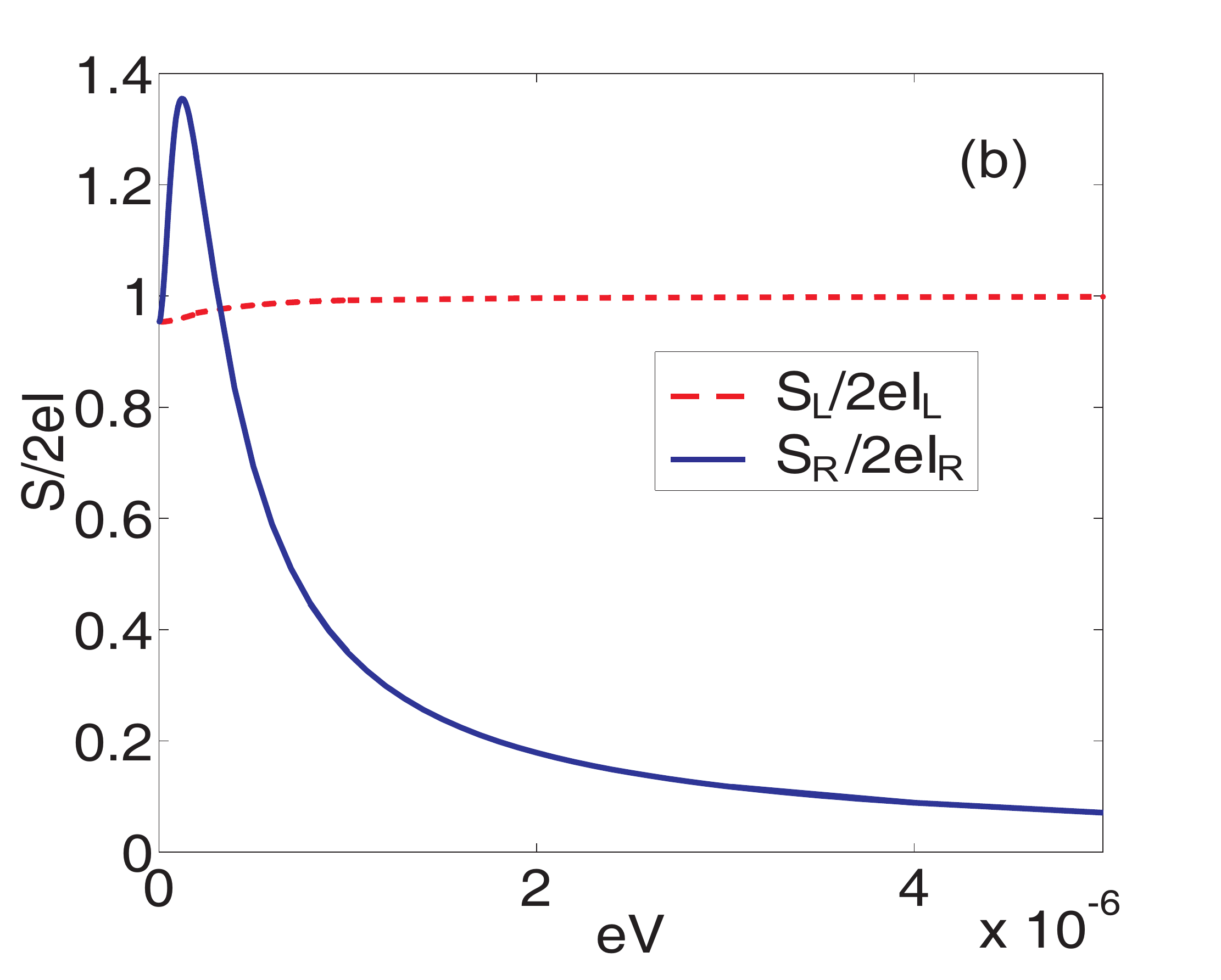}
	\caption{\label{4} (a) The bias-voltage dependence of the conductances (a) and the Fano factors (b) of the leads. Inset of (a): the zero-bias behavior of the conductances of the half-metallic (solid and dashed) and paramagnetic (dotted) leads. Parameters correspond to the point A in Figure \ref{3}a ($V_{x}=0.95$, $V_{z}=0.8$).}
\end{figure*}
In order to explain given magnetic-field orientation dependencies in more details it is useful to consider two limit cases: perpendicular $V_{1}\left(V_{x}=0,~V_{z}=1\right)$ and longitudinal $V_{2}\left(V_{x}=1,~V_{z}=0\right)$ magnetic fields. In $V_{1}$ the z-component of the MBS spin polarization $P_{M,i}=P_{M,i\uparrow}-P_{M,i\downarrow}$ is negative at both edges and $P_{M,L}=P_{M,R}$ whereas $P_{M,R}=-P_{M,L}$ in $V_{2}$. Hence, there is a range of the magnetic-field angles close to $\pi/4$ where $P_{M,R}\approx0$ \cite{valkov-16a,valkov-17b}. At the same time $P_{M,L}$ for this particular direction has a maximal amplitude since on-site energy splitting is originated from both $V_{x}$ and $V_{z}$ components \cite{kwapinski-17}. Similar situation is observed at the other magnetic-field orientations: in each quarter of the canted field the absolute value of the z-component of the MBS spin polarization is maximal at one end and minimal at the opposite. Note that our qualitative formula of $P_{M,i}$ gives the numerical results which just quantitatively differ from the ones based on the rigorous definition \cite{sticlet-12}.

\section{\label{sec5} Magnetic-field orientation dependence of the conductance and shot noise}

Now we turn to the behavior of the MBS-assisted conductance and noise. In spin-polarized regime the transport is strongly determined by not just pure coupling strengths $\Gamma^{+,-}_{L\left(R\right)}$ but rather the effective ones which also include the information about the spin-dependent lifetime of the MBSs proportional to $P_{M,L\left(R\right)\sigma}$. This is confirmed by the magnetic-field orientation dependencies of the low-bias ($eV=10^{-4}$) conductances of the left and right leads in Figures \ref{3}a and b respectively. Here we start with the situation where the magnetizations are parallel and $\sigma=\uparrow$ is a majority-spin projection. Both maps look similar as the sequences of concentric rings where the conductance is about $2G_0$ (hereinafter the MBS rings). Such peculiarities are the result of resonant transport via the MBSs \cite{flensberg-10,law-09}. The periodical appearance of the conductance maximum with increasing magnetic-field amplitude is in agreement with the oscillations of the MBS energy \cite{valkov-16a} caused by the magnetic-field periodical dependence of the coupling between the MBSs, $t_{\gamma}$ \cite{dassarma-12,lim-12,rainis-13}. Since there are only spin-up carriers in the leads the effective couplings with the wire and the conductances are greater in the lower half-plane, according to the model Hamiltonian (\ref{HW}). It is explained by the above-observed properties of the spin-up MBS probability densities (see Figs.\ref{2}a,c) and nonlocal character of the transport. In general, the conductance of the contact with topologically SC system is greater if the lead's polarization is parallel to the magnetic field that is confirmed by experiment \cite{sun-16}. In the upper half-plane the MBS rings of $G_L$ and $G_R$ become thinner and the breaks appear for two magnetic-field orientations, $V_{x} \approx V_{z}$ and $V_{x} \approx -V_{z}$, respectively. Thus, we get obvious current-symmetry breaking, $I_{L}\neq-I_{R}$, in these parametric regions. Such a situation in charge nonconserving system is not realized in the symmetric coupling and voltage regime, $\Gamma_{L}=\Gamma_{R}$ and $V_{L}=V_{R}$, if leads are paramagnetic \cite{dasilva-14,you-15}.

\begin{figure*}[htbp]
    \includegraphics[width=0.4\textwidth]{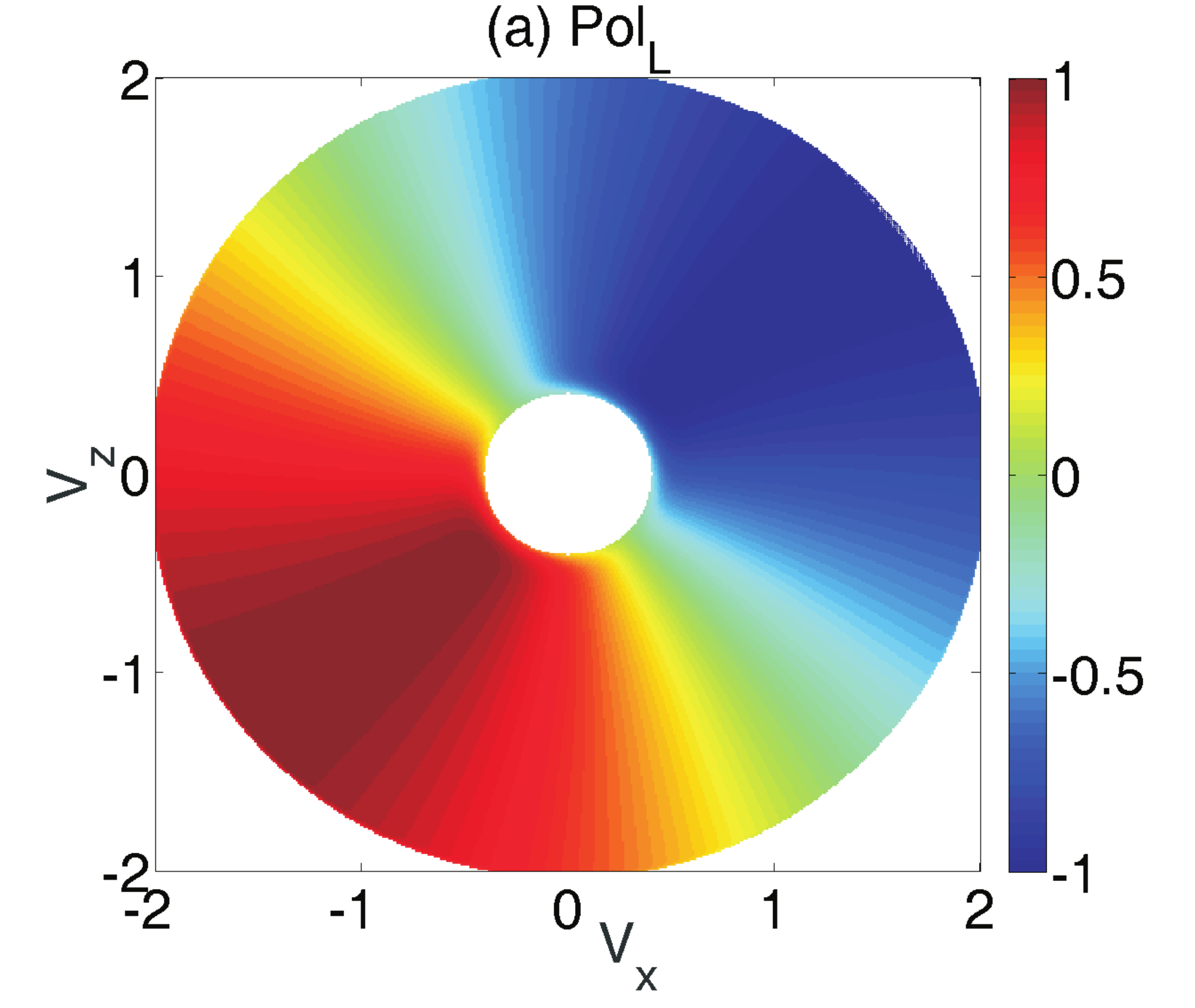}
    \includegraphics[width=0.4\textwidth]{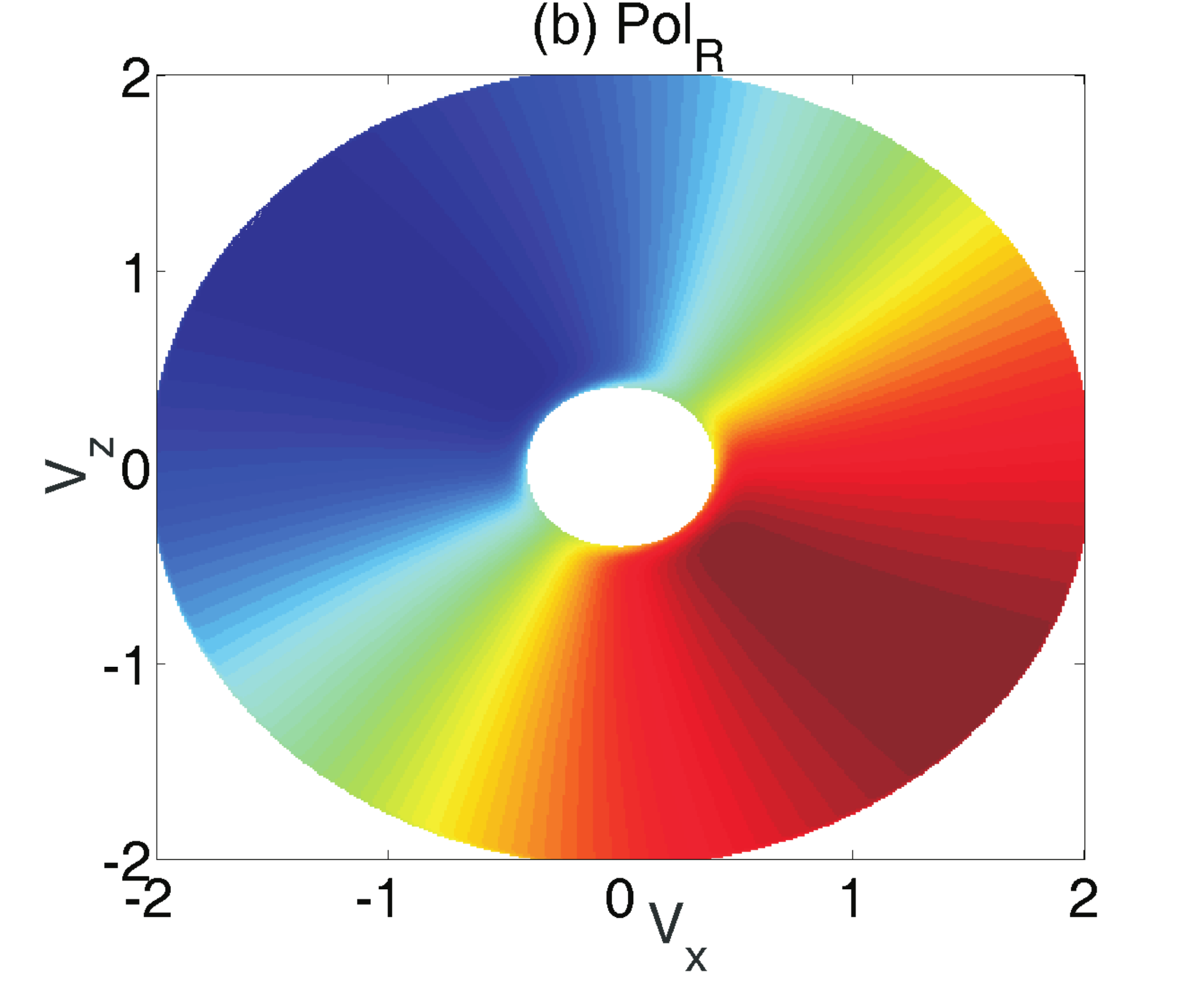}
    \caption{\label{6} The magnetic-field orientation dependence of the spin polarization of the current in the left (a) and right (b) paramagnetic lead. Parameters are the same as in Figure \ref{2}.}
\end{figure*}
In our system perfectly spin-polarized current from the left (right) lead is significantly hampered due to the vanishing same-spin MBS probability density at the left (right) wire's end. As a result, the left (right) effective coupling strength is extremely decreased and the regime turns to be strongly asymmetric at $V_{x} \approx V_{z}$ ($V_{x} \approx -V_{z}$). Consequently, at both orientations the transport properties of the system are qualitatively similar to those observed at one-lead geometry. For example, if $V_{x} \approx V_{z}$ the wire is virtually connected only with the right lead. Then $G_{L}$ is close to zero and $G_{R}$ tends to unity. To explain it minutely we plotted the left- and right conductances as functions of bias voltage in the magnetic field corresponding to the Majorana zero mode (see the point A in Fig. \ref{3}a). It is seen that the right conductance is $2G_0$ at low voltages but at zero bias it forms a sharp dip (see Fig. \ref{4}a). That is similar to the behavior of the MBS-assisted one-lead transport \cite{flensberg-10,wu-12}. However, the weak but non-zero coupling with the left lead gives rise to $G_{L,R}\left(eV=0\right)\neq0$ (see inset of Fig. \ref{4}a). It is also in agreement with the predictions of the model-independent scattering matrix theory valid in the low-energy limit \cite{nillson-08}. In particular, it explains the influence of $t_{\gamma}$ and $\Gamma_{L,R}$ on the conductance peak position. In extremely asymmetric situation the ZBP splitting and the local minimum at the zero bias emerge when $t_{\gamma}$ exceeds even very small critical value \cite{wu-12}.

The  left- and right low-bias current-noise calculations display asymmetry as well (see Fig. \ref{3}c,d). The right Fano factor, $F_{R}=S_{R}/2eI_{R}$, approaches 2 at low conductance if $V_{x} \approx V_{z}$ (see the red noise trail at Fig. \ref{3}d). One points out that the transport in this case is predominantly mediated by equal-spin local AR \cite{law-09,he-14} and is typical for one-lead regime. In opposite, the left Fano factor, $F_{L}$, approaches 1 for all the magnetic-field amplitudes as the coupling with the SC wire is suppressed for the given orientation. The Fano factors corresponding to the MBS rings are zero indicating the Majorana-induced resonant Andreev tunneling \cite{wu-12}. The bias-voltage dependence of the Fano factors for $V_{x} \approx V_{z}$ at the Majorana zero mode additionally shows that at zero bias the perfect analogy with the one-lead case cannot be drawn since $F_{L,R}\simeq1$ (see Fig.{4}b). In addition, the right noise in the red-trail region of Fig. \ref{3}d ($eV/2<E_{M}$) also tends to unity at zero bias (not shown at Fig. \ref{4}b). Those are the signatures of the strongly asymmetric two-lead transport \cite{nillson-08,wu-12} (see also \ref{apx1}). If the leads are paramagnetic there is no conductance asymmetry, $G_{L}=G_{R}$, and the ZBP is observed (see dotted line in the inset of Fig.{4}a).

\begin{figure}[htbp]
	\includegraphics[width=0.4\textwidth]{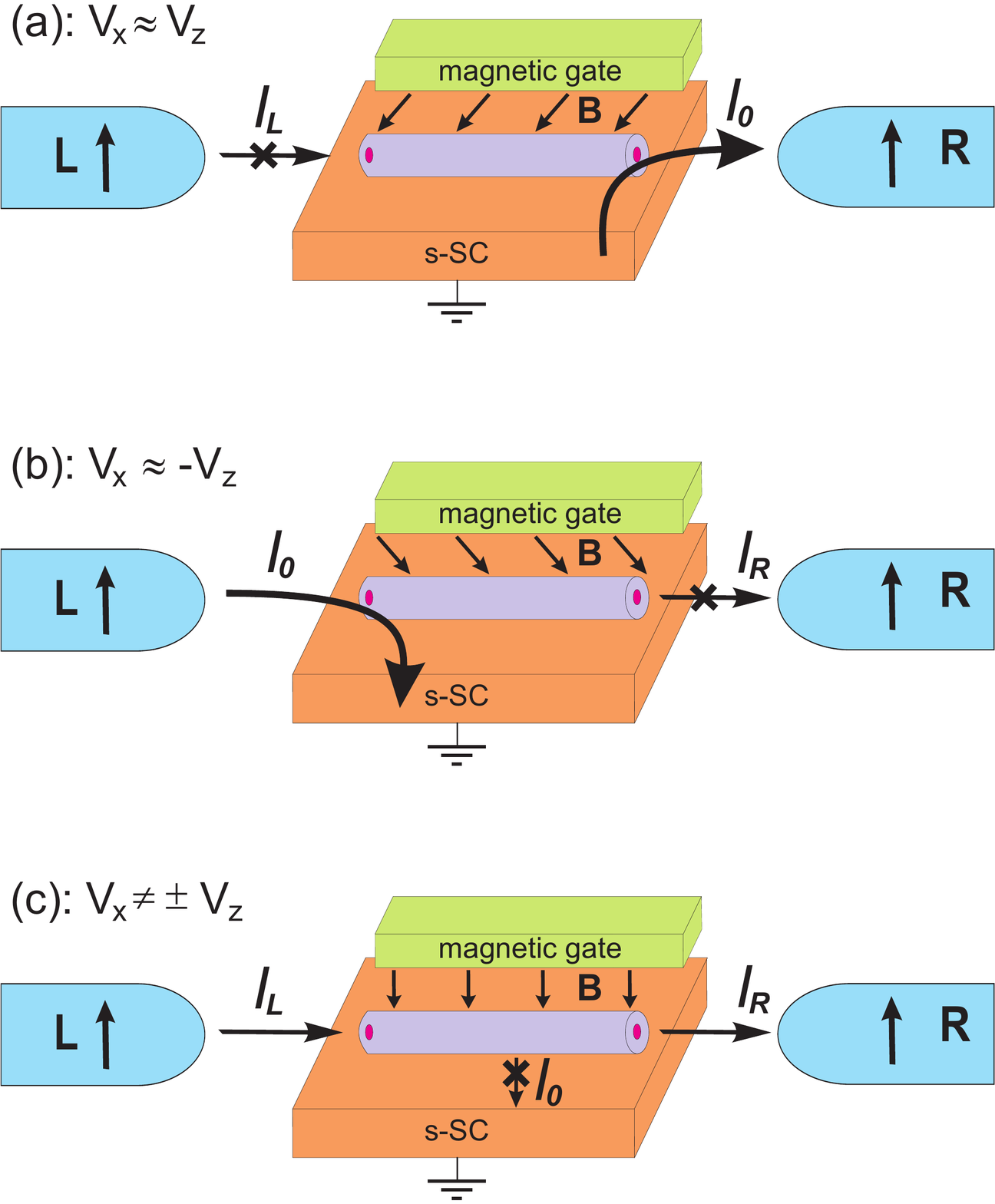}
	\caption{\label{switch} The scheme of the MBS-assistant current switch. (a) $I_L \rightarrow 0$ and $I_0 \approx -I_R$ since $P_{M,L}=P_{M,L\downarrow}$ and $\theta_L=0$ for $V_{x}>0,~V_z \approx V_x$; (b) $I_R \rightarrow 0$ and  $I_0 \approx -I_L$ since $P_{M,R}=P_{M,R\downarrow}$ and $\theta_R=0$ for $V_{x}<0,~V_z \approx -V_x$; (c) $I_0 \rightarrow 0$ and $I_{L} \approx -I_{R}$ if $V_{x} \neq\pm V_{z}$.}
\end{figure}
Taking into account the behavior of the MBS spin-resolved probability densities it is not difficult to describe the dependencies of the conductances and Fano factors analogous to those in Figures \ref{3} if there is antiparallel configuration of the half-metallic leads ($\theta_{L}=0,\theta_{R}=\pi$). In this case the MBS-assisted spin-polarized transport is substantially defined by the densities $P_{M,L\uparrow}$ and $P_{M,R\downarrow}$. Then, according to Figures \ref{2}a and d, the width of the MBS rings is greater in the half-plane $V_{x}<0$. In turn the gaps in the rings appear in $V_{x}>0$: $V_{z}\approx V_{x}$ for $G_{L}$ and $V_{z}\approx -V_{x}$ for $G_{R}$. Consequently, the tunnel magnetoresistance (TMR) of both leads, $TMR_{L,R}=\frac{G_{L,R}^{P}-G_{L,R}^{AP}}{G_{L,R}^{AP}}$, significantly increases for the magnetic fields $V_{x}>0, V_{z}<0$. And it takes extremely high values if $V_{x}>0,~V_{z}\approx -V_{x}$. TMR is exactly zero in the MBS rings.

\begin{figure}[htbp]
	\includegraphics[width=0.45\textwidth]{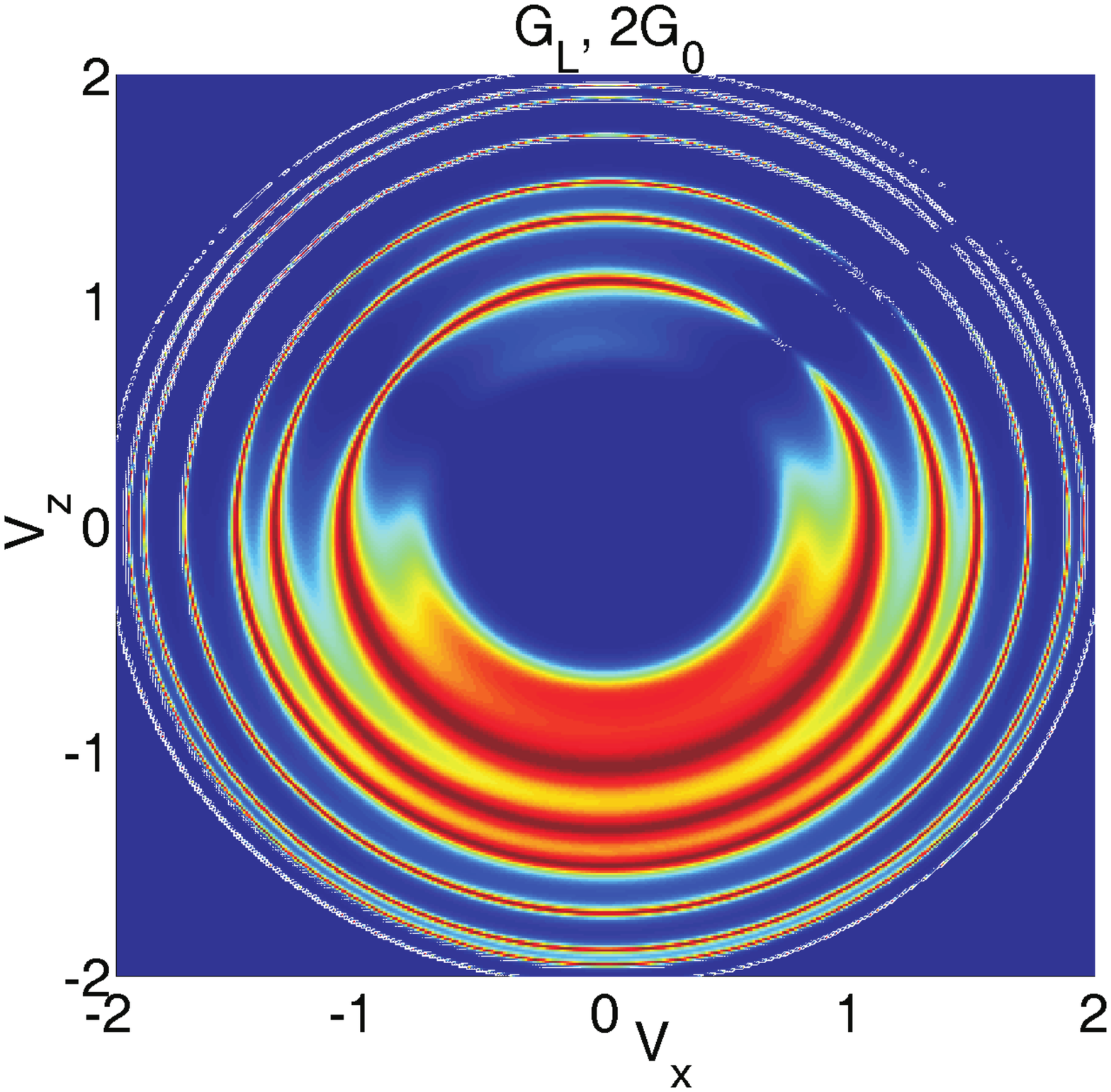}
	\caption{\label{imp} The magnetic-field orientation dependence of the left conductance in the presence of Anderson disorder in the SC wire. Parameters are the same as in Figure \ref{2}.}
\end{figure}
\begin{figure*}[htbp]
	\includegraphics[width=0.5\textwidth]{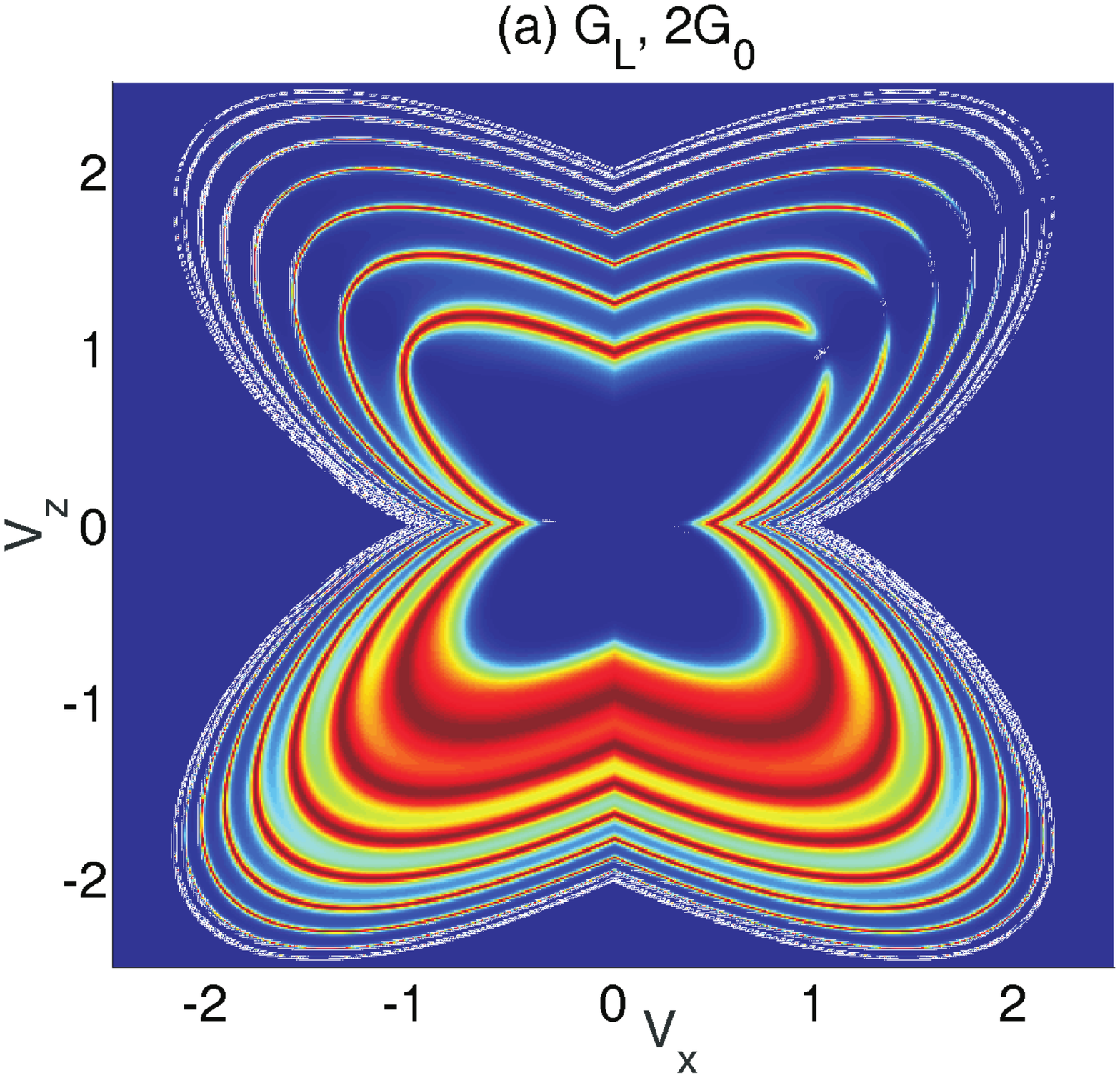}
	\includegraphics[width=0.5\textwidth]{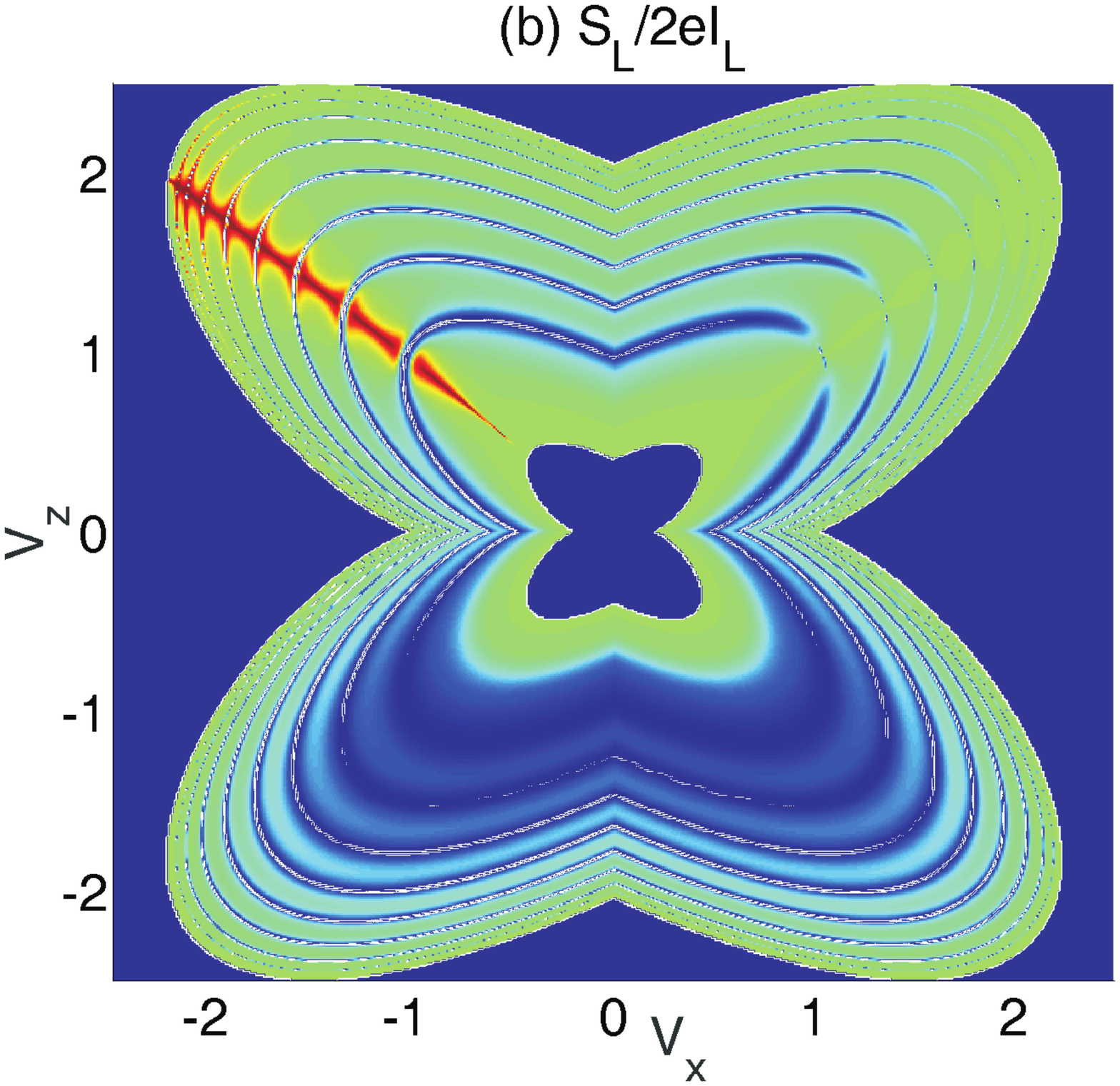}
	\caption{\label{ani} The magnetic-field orientation dependence of the left-lead conductance (a) and Fano factor (b) taking into account g-factor anisotropy. Parameters are the same as in Figure \ref{2}.}
\end{figure*}
\begin{figure}[htbp]
	\includegraphics[width=0.4\textwidth]{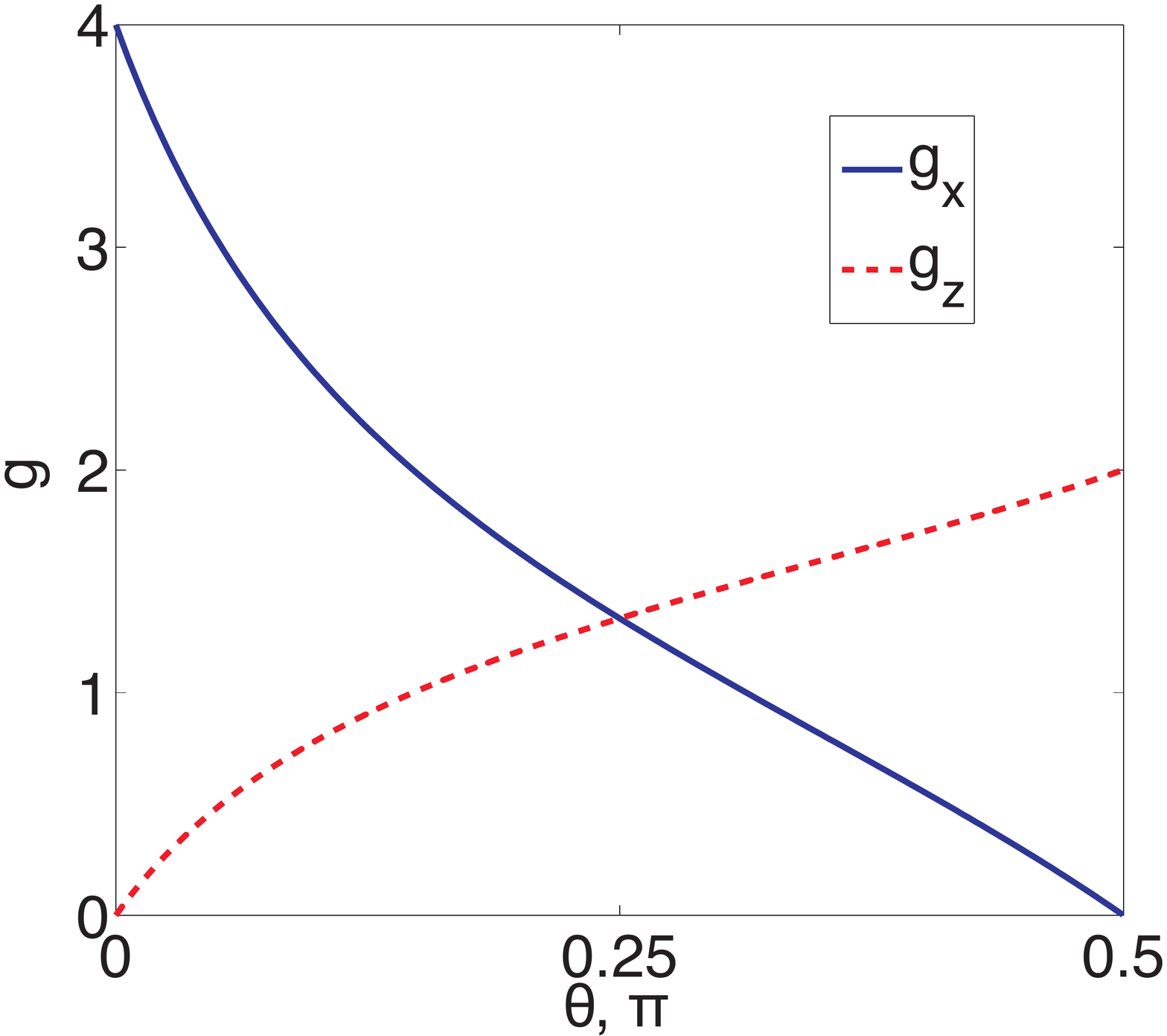}
	\caption{\label{g_ani} The magnetic-field orientation dependencies of $g_x$ and $g_z$.}
\end{figure}
The effect of the magnetic field on the spin-polarized transport
in the topologically SC wire is similar to the effect of the
electric gate field in the spin transistor described by Datta and
Das \cite{datta-90}. Thus, the SC wire can be considered as an
active element of spin field-effect transistor if the magnetic
field in the system is controlled by the magnetic gate.
Moreover, the obtained nontrivial MBS spin
polarization can be exploited to obtain perfect spin-filtering
effect in the situation with paramagnetic leads. In Figures
\ref{6}a and b the corresponding spin polarization,
$Pol_{i}=\frac{G_{i\uparrow}-G_{i\downarrow}}{G_{i\uparrow}+G_{i\downarrow}}$,
is plotted. The behavior of $Pol_{i}$ is similar to $P_{M,i}$.
There are four magnetic-field orientations where the current in
one lead is perfectly spin-polarized but absolutely unpolarized in
the opposite lead.

It is worth to note that from experimental point of view the
conductance of the SC wire,
$G=\frac{1}{2}\frac{d}{dV}\left(I_{L}-I_{R}\right)=\frac{1}{2}\left(G_{L}+G_{R}\right)$,
is more interesting. Taking into account the behavior of $G_{L}$
and $G_{R}$ (Fig.\ref{3}a,b) we see that the height of the MBS rings of $G$ for two magnetic-field
orientations (with approximately $\pi/2$ angle between them) is  $G_{0}$ instead
of 2$G_{0}$. In other words the direction of the current in the SC
substrate, $-I_0=I_L+I_R$, can be controlled by the magnetic field. If $V_z \approx V_x$ and $eV/2 > E_{M}$ that  $I_L \approx 0$ (since in the low-bias regime only the MBS participates at transport), $I_0 \approx \mid I_R \mid$ as it is shown in Fig. \ref{switch}a. Similarly, if $V_z \approx
-V_x$ and $eV/2 > E_{M}$ that $I_R \approx 0$, $I_0 \approx -\mid I_L \mid$ (Fig. \ref{switch}b). Finally, if $V_{x} \neq\pm V_{z}$ and $eV/2 > E_{M}$ that $I_{L} \approx -I_{R}$ and $I_0 \approx 0$ (Fig. \ref{switch}c). In contrast to \cite{cao-12} the conductance corresponding to $I_0$ is not zero  for $V_z \approx
\pm V_x$ if the coupling of the SC wire with the
leads is symmetric, $\Gamma_L=\Gamma_R$, and is approximately
equal to $2G_0$. Thus, the effect of the current switch can be used
to detect Majorana fermions and in electronic applications. In \ref{apx2} the influence of left and right magnetization orientations on the current asymmetry is presented.

\section{\label{sec6}Influence of disorder and g-factor anisotropy}

In this section we discuss the impact of some factors which arise in experimental situations on the observed features of the transport properties. First is the presence of disorder due to impurities in the wire. We considered the influence of Anderson disorder, modeled as an additional on-site random potential $w_j$ varying in range $\left[-t/2,~t/2\right]$, on the transport. The left-lead conductance is depicted in Fig. \ref{imp}. The magnetic-field orientation dependence is not changed qualitatively. However, the width of the MBS rings becomes thinner. And, as a result, the regions where the MBS rings are broken become slightly wider.

\begin{figure*}[htbp]
	\includegraphics[width=0.4\textwidth]{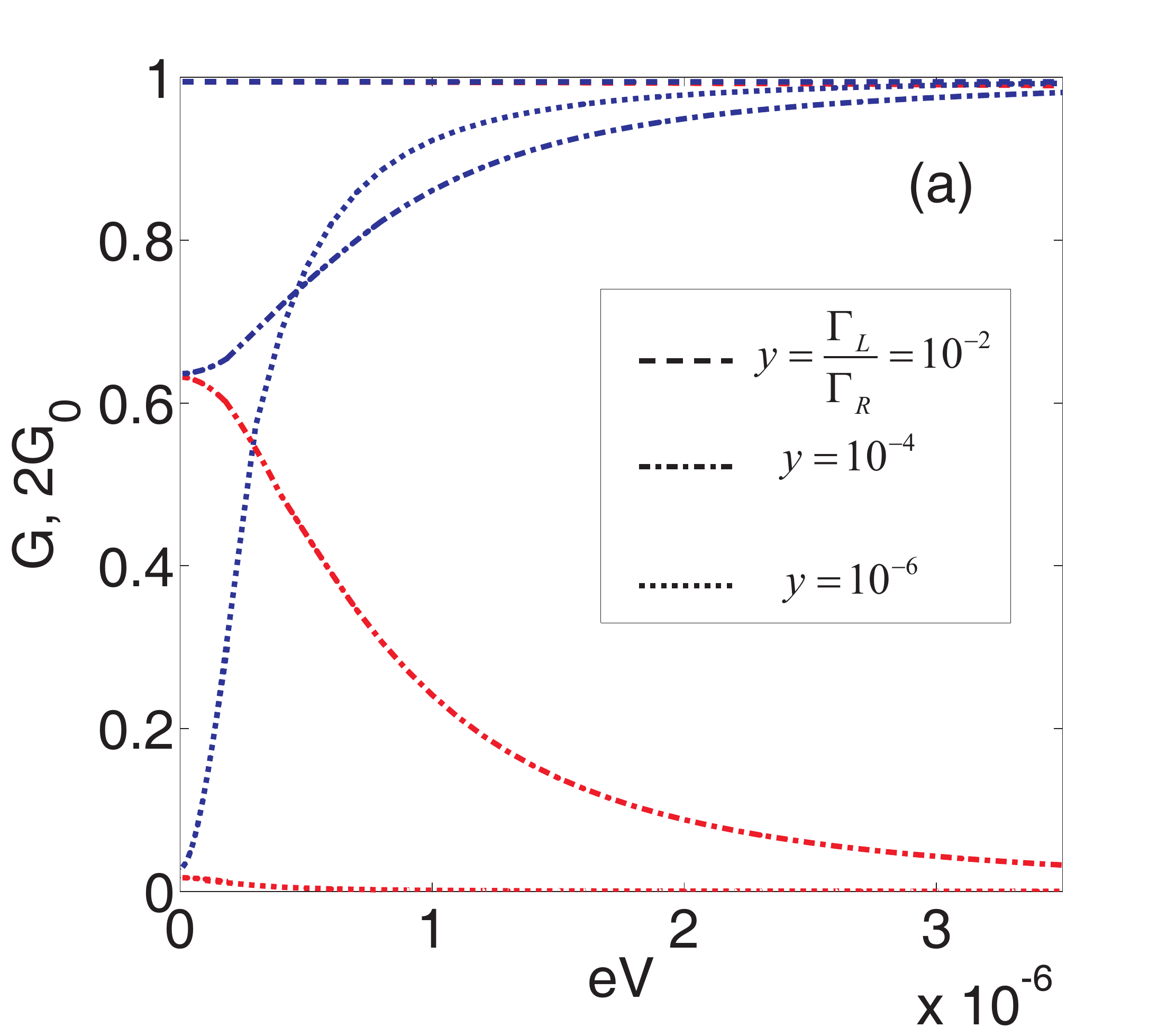}
	\includegraphics[width=0.4\textwidth]{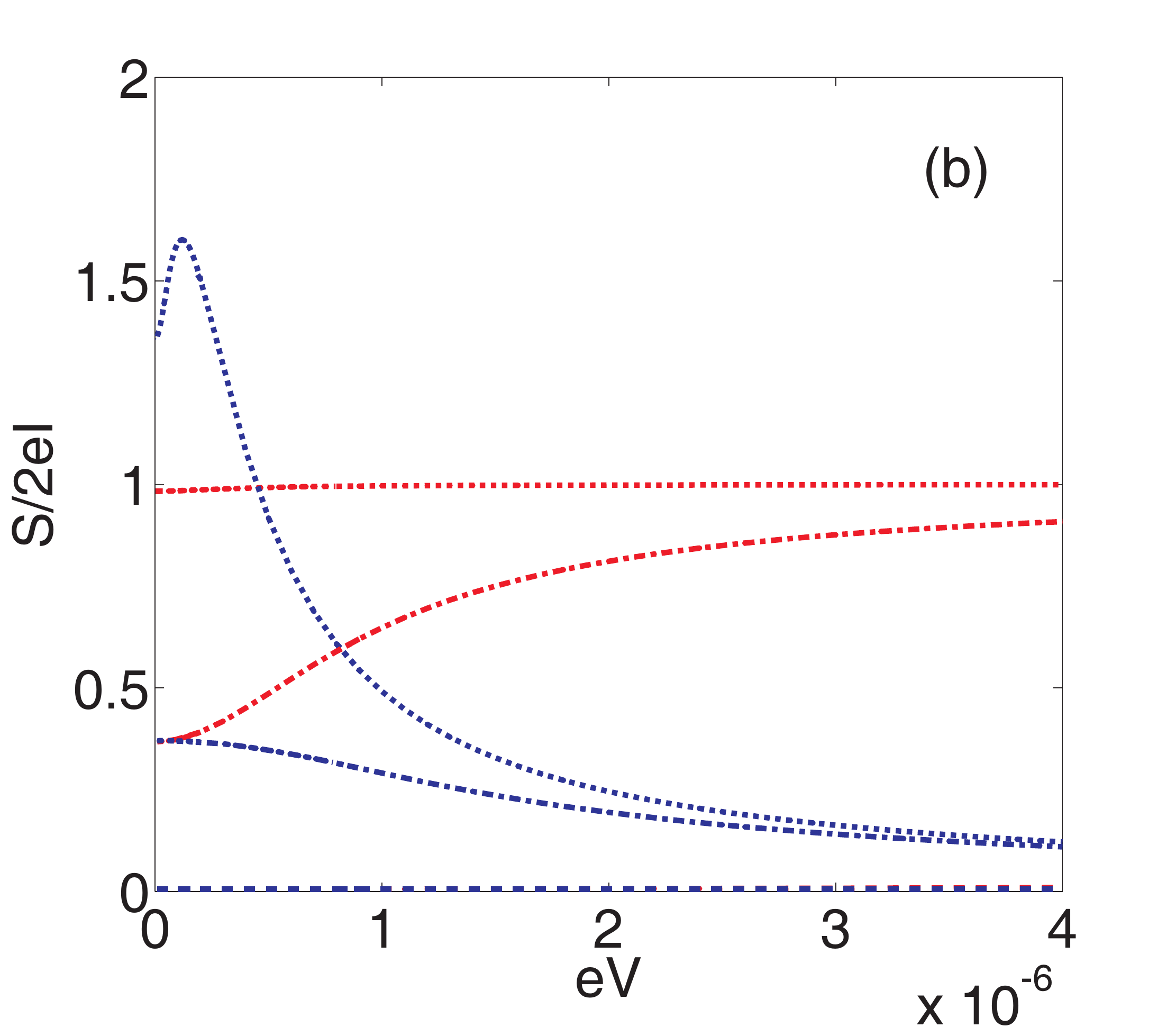}
	\caption{\label{5} The bias-voltage dependence of the conductance (a) and noise Fano factor (b) of the paramagnetic leads for different asymmetry parameter $y=\frac{\Gamma_L}{\Gamma_R}$. Parameters correspond to the point A in Figure \ref{3}a.}
\end{figure*}
Recently considerable attention has been paid to the problems of g-factor asymmetry in semiconducting low-dimensional structures. The experiments revealed significant change of the g-factor when magnetic field was rotated \cite{lee-14,qu-16}. In semiconducting nanowires which are perspective to experience topologically nontrivial phase orbital motion of carriers and the existence of subbands due to confinement are considered as a mechanism responsible for the g-factor anisotropy \cite{maier-14,winkler-17}. Since the studied structure is exactly one-dimensional we used a phenomenological model based on the theoretical \cite{winkler-17} and experimental \cite{lee-14} results. Specifically, it was supposed that g-factor decreases two times when the magnetic field is rotated by $\pi/2$, i.e. $g_x=4$ and $g_z=2$. The derived magnetic-field orientation dependencies of $g_x$ and $g_z$ are plotted in Fig. \ref{g_ani}. The resulting surfaces of the left-lead conductance and Fano factor experience transformation because of new topological phase conditions, $\mu^2+\Delta^2<\frac{1}{4}\left(g_{x}^2V_{x}^2+g_{z}^2V_{z}^2\right)<\left(2t - \mu\right)^2+\Delta^2$ (see Figs. \ref{ani}a and b respectively). It leads to the dependence of the Majorana zero mode on the magnetic-field angle which is clearly seen in Figs. \ref{ani}a,b. However, the above-described features are not changed qualitatively. In particular, the MBS rings become flower-shape dependencies where the MBS resonances are suppressed at $\theta \approx \frac{\pi}{4}$. The Fano-factor red trail remains at $\theta \approx \frac{\pi}{4}+\frac{\pi}{2}$.

Thus, the suppression of the MBS rings and, consequently, current-switch effect can persist under the presence of diagonal disorder and g-factor anisotropy.

\section{\label{sec7}Conclusion}

\begin{figure*}[htbp]
	\includegraphics[width=0.4\textwidth]{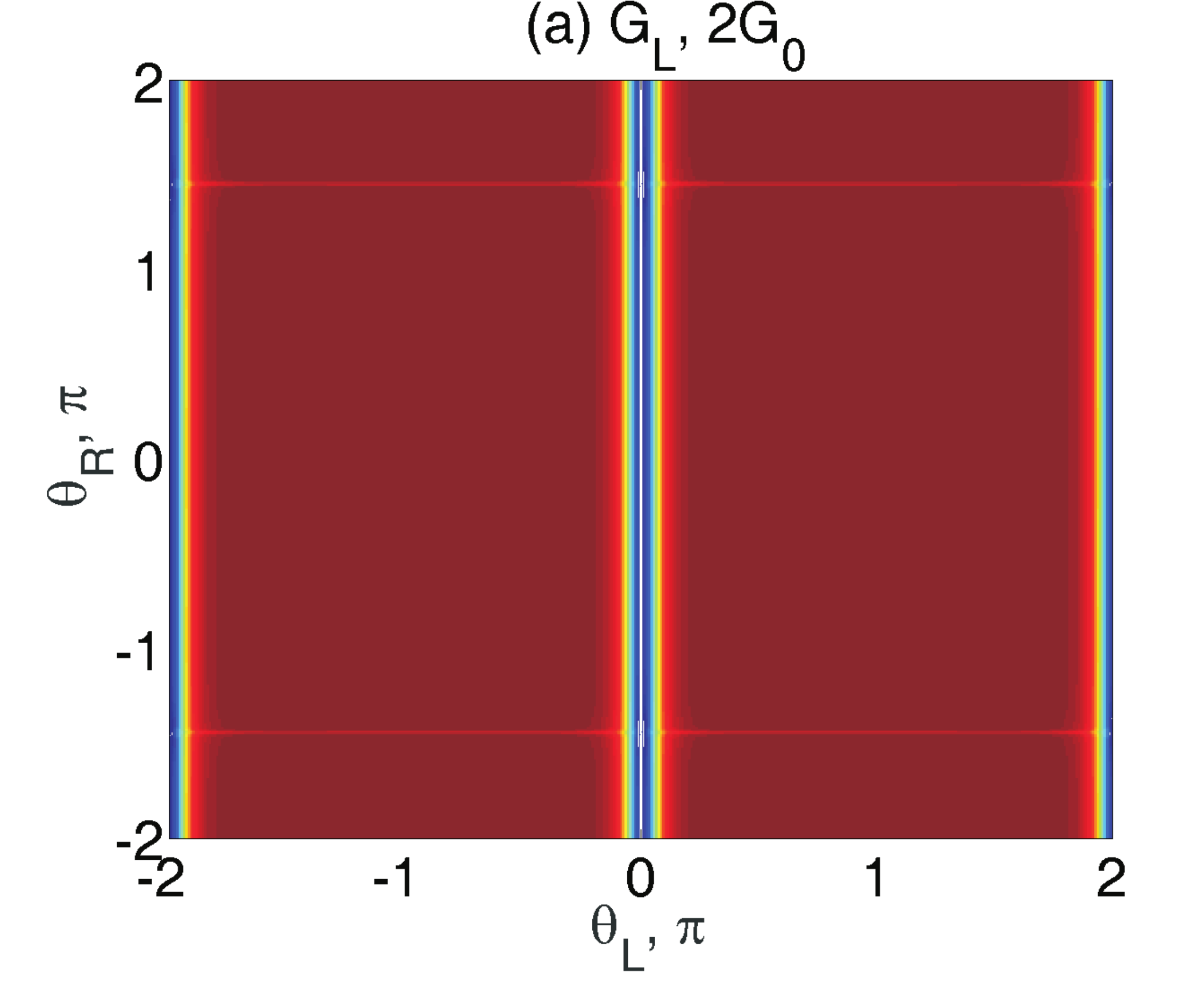}
	\includegraphics[width=0.4\textwidth]{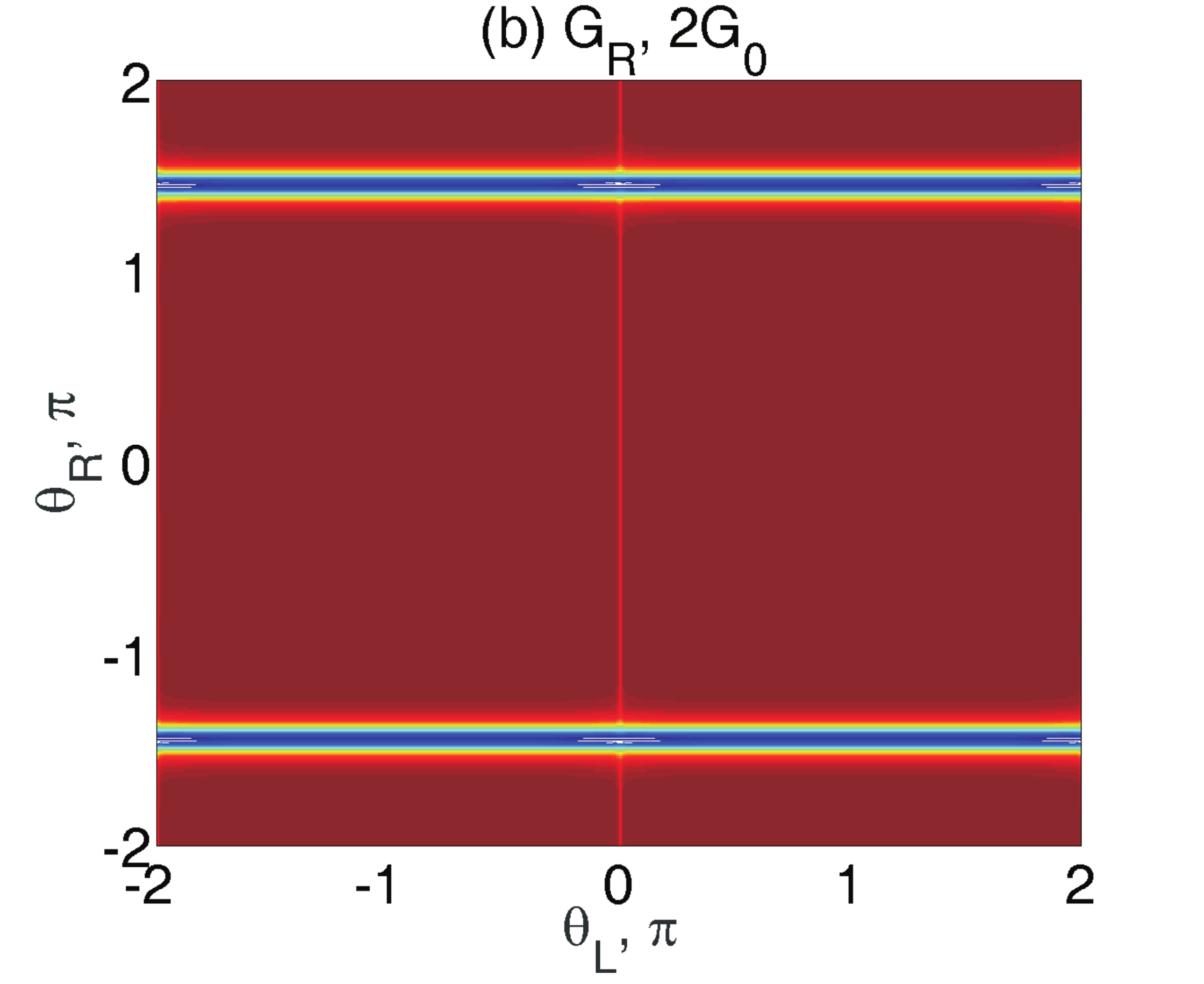}
	\caption{\label{7} The magnetization-angle dependencies of the (a) left- and (b) right conductances. Parameters correspond to the point A in Figure \ref{3}a.}
\end{figure*}
\begin{figure*}[htbp]
	\includegraphics[width=0.4\textwidth]{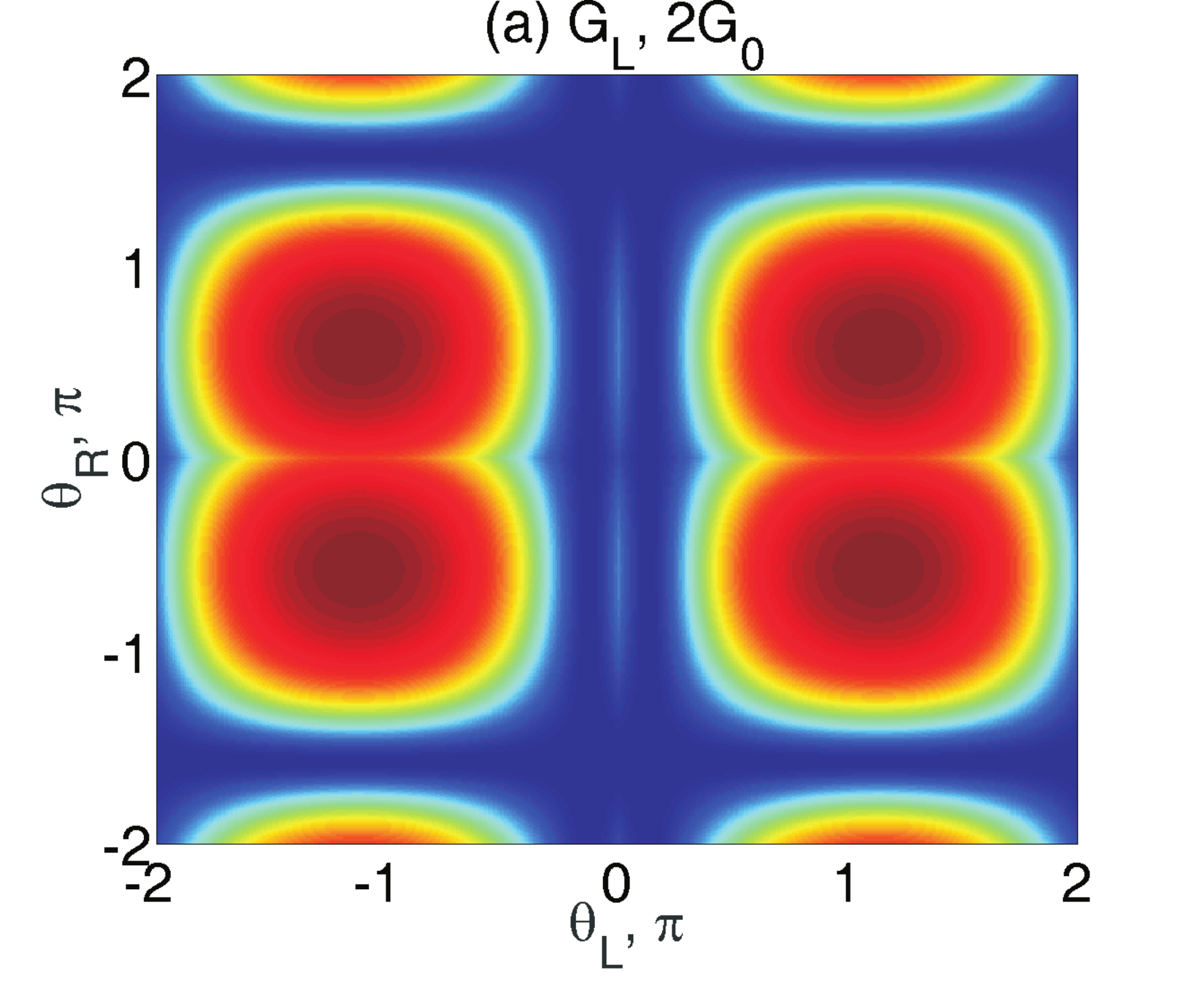}
	\includegraphics[width=0.4\textwidth]{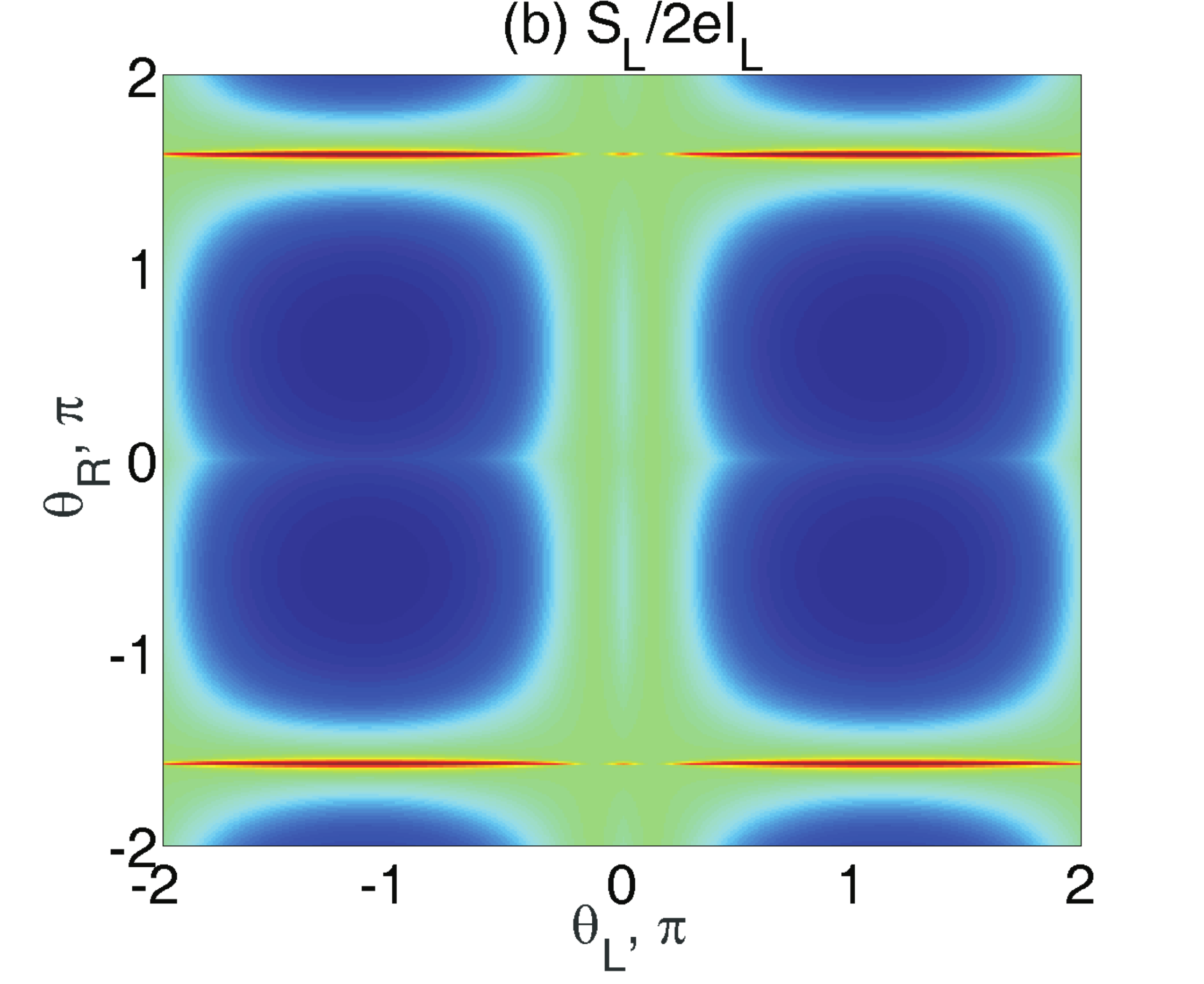}
	\caption{\label{8} The magnetization-angle dependencies of the left conductance (a) and Fano factor (b). Parameters correspond to the point B in Figure \ref{3}a.}
\end{figure*}
We studied the influence of the magnetic-field orientation on quantum transport in the SC wire. Using the nonequilibrium Green's functions and microscopic tight-binding description of the SC wire it is demonstrated that the low-bias conductance and shot noise behavior is defined by the MBS spin polarization if the leads are half-metals. In particular, the width of the MBS rings in the lead conductance increases for the magnetic-field orientations in which the lead magnetization and the MBS spin polarization are oriented in the same direction. If they are antiparallel the corresponding conductance vanishes. Simultaneously, the MBS rings in the conductance of opposite lead survive and the corresponding Fano factor displays specific noise trail indicating the domination of the MBS-assisted local AR processes. These effects give rise to strong current asymmetry in the system. Taking it into account we propose the MBS-assisted current switch device where charge carrier flow can be controlled by magnetic gate. There are three possible paths of the current: (a) from the substrate to the right lead if $V_{x}\approx V_{z}$; (b) from the left lead to the substrate if $V_{x}\approx -V_{z}$; (c) from the left to right lead if $V_{x}\neq \pm V_{z}$. This feature survives in the presence of Anderson-type disorder and phenomenologically modeled g-factor anisotropy. The control of the spin polarization of the current in paramagnetic leads by magnetic field is shown. In addition, we demonstrated the possibility to manipulate the MBS-assisted current by changing the magnetization direction. The obtained features of the transport properties can be used to detect the MBSs in spin-polarized spectroscopy/microscopy experiments and develop electronic and spintronic applications based on the topologically SC wires.

\ack
We acknowledge fruitful discussions with A.D. Fedoseev and M.S. Shustin. This work was financially supported by the Comprehensive programme SB RAS no. 0356-2015-0403, Russian Foundation for Basic Research, Government of Krasnoyarsk Territory, Krasnoyarsk Region Science and Technology Support Fund to the research projects nos. 16-02-00073, 16-42-243056, 16-42-242036, 17-42-240441. The work of A.S.V. was supported by grant of the President of the Russian Federation (MK-1398.2017.2).

\appendix

\section{\label{apx1}Superconducting wire between paramagnetic leads in asymmetric regime}

The bias-voltage dependencies of the conductance and noise similar to the above-described in Fig. \ref{4} can be observed when $y=\frac{\Gamma_L}{\Gamma_R}\neq1$ \cite{wu-12,dasilva-14,you-15}. It is clearly seen in Figures \ref{5} that the decreasing of $y$ completely changes the conductances. If $y=10^{-2}$ the asymmetry in the transport characteristics can be still neglected. However, for $y=10^{-6}$ we get the curves resembling the ones in Fig. \ref{4}a. The Fano factor of the lead stronger coupled with the topologically SC wire tends to $2$ at low and zero bias if $y$ diminishes (see the set of blue curves in Fig. \ref{5}b) pointing out the MBS-assisted local AR. On the contrary, the noise characteristic of the opposite end approaches unity. Note that in the situation of paramagnetic or unparallel half-metallic leads crossed AR additionally contributes to nonlocal transport \cite{nillson-08} and is equal to the elastic cotunneling part in the linear response regime \cite{law-14,floser-13,liu-13}.

\section{\label{apx2}Influence of the orientations of the magnetizations on the conductance and noise}

In \cite{wu-14,chen-16} the significant influence of the orientations of the molecular fields in the ferromagnetic leads, $\theta_{L,R}$, on the MBS-mediated AR and $t_{\gamma}$ was investigated. In this Appendix we show how the above-described current asymmetry depends on $\theta_{L,R}$. For the magnetic field fitting the case $G_{L}=0$, $G_{R}=1$ and $\theta_{L,R}=0$ (see the point A in Fig. \ref{3}a) the magnetization-angle dependencies of the conductances are plotted in Figures \ref{7}. It is seen in Fig. \ref{7}a that the $G_L$ minimum are not affected by the $\theta_{R}$. It emphasizes the local nature of this conductance suppression when at one edge the magnetization and the MBS spin polarization become oppositely directed. The conductance $G_R$ can also be suppressed for certain $\mathbf{h}_{R}$ directions (see the horizontal areas in Fig. \ref{7}b). Moreover, there are two configurations of the magnetizations ($\theta_L=0,~\theta_R\approx\frac{3}{2}\pi$) and ($\theta_L=0,~\theta_R\approx-\frac{3}{2}\pi$) when both $G_{L}$ and $G_{R}$ vanish.

The conductance behavior may be more sophisticated. To demonstrate the different possible scenarios the other initially low-conductance point is considered (the point B in Fig. \ref{3}a). In Fig. \ref{8}a the left conductance is presented (the map of $G_R$ is very similar in this case). The dependence is the combination of previously obtained ones. The two couples of the regions where $G_L$ is suppressed are well-defined. In each couple the areas are located symmetrically with respect to $\theta_{L}$ or $\theta_{R}$ axis. Simultaneously, the shot noise plot gives additional information (see Fig. \ref{8}b). It points out that the vertical areas are caused by tending to zero left effective tunnel coupling since $F_{L}\rightarrow1$ and $F_{R}\rightarrow2$ (the last is not shown here). It is important to note that as in the Fig. \ref{7}a the angle of right magnetization doesn't influence on the left conductance in these zones. The horizontal areas are characterized by $F_{L}\rightarrow2$ and $F_{R}\rightarrow1$ and can be explained by the magnetization-angle impact on $t_{\gamma}$ \cite{chen-16}. Thus, for some orientations of the magnetizations the current-switch effect can be suppressed.

\section*{References}

\bibliography{VV_AS_arxiv}

\end{document}